\documentclass[preprints,article,accept,moreauthors,pdftex]{Definitions/mdpi}
\firstpage{1} 
\pubvolume{xx}
\issuenum{1}
\articlenumber{5}
\pubyear{2019}
\copyrightyear{2019}
\history{Received: date; Accepted: date; Published: date}
\pdfoutput=1

\Title{On Geometry of Information Flow for Causal Inference}

\Author{Sudam Surasinghe $^{1}$ and Erik M.~Bollt $^{2}$}

\AuthorNames{Sudam Surasinghe, Ioannis Kevrekidis and Erik M.~Bollt}

\address{%
$^{1}$ \quad Department of Mathematics, Clarkson University, Potsdam, NY 13699\\
$^{2}$ \quad Electrical and Computer Engineering and $C^3S^2$
the Clarkson Center for Complex Systems Science, Clarkson University, Potsdam, New York 13699}

\abstract{Causal inference is perhaps one of the most fundamental concepts in science, beginning originally from the works of some of the ancient philosophers, through today, but also weaved strongly in current work from statisticians, machine learning experts, and scientists from many other fields.  This paper takes the perspective of information flow, which includes the Nobel prize winning work on Granger-causality, and the recently highly popular transfer entropy, these being probabilistic in nature. Our main contribution will be to develop analysis tools that will allow a geometric interpretation of information flow as a causal inference indicated by positive transfer entropy. We will describe the effective dimensionality of an underlying manifold as projected into the outcome space that summarizes information flow.  Therefore contrasting the probabilistic and geometric perspectives, we will introduce a new measure of causal inference based on the fractal correlation dimension conditionally applied to competing explanations of future forecasts, which we will write  \texorpdfstring{$GeoC_{y\rightarrow x}$}{Lg}.  This avoids some of the boundedness issues that we show exist for the transfer entropy, \texorpdfstring{$T_{y\rightarrow x}$}{Lg}. 
We will highlight our discussions with data developed from synthetic models of successively more complex nature: these include the H\'{e}non map example, and finally a real physiological example relating breathing and heart rate function.}


\begin{document}
%%%%%%%%%%%%%%%%%%%%%%%%%%%%%%%%%%%%%%%%%%

%%%%%%%%%%%%%%%%%%%%%%%%%%%%%%%%%%%%%%%%%%
\section{Introduction}

Causation Inference is perhaps one of the most fundamental concepts in science, underlying questions such as ``what are the causes of changes in observed variables."  Identifying, indeed even defining causal variables of a given observed variable is not an easy task, and these questions date back to the Greeks \cite{Williams1973-WILAPB,sep-aristotle-causality}.  This includes important contributions from more recent luminaries such as Russel \cite{Russel2015}, and from philosophy, mathematics, probability, information theory, and computer science. We have written that, \cite{OpenVcloseEr2018}, ``a basic question when defining the concept of information flow is to contrast versions of reality for a dynamical system. Either a subcomponent is closed or alternatively there is an outside influence due to another component."  Claude Granger's Nobel prize \cite{NobelGren2004} winning work leading to Granger Causality (see also Wiener, \cite{Wiener1956}) formulates causal inference as a concept of quality of forecasts. That is, we ask, does system $X$ provide sufficient information regarding forecasts of future states of system $X$ or are there improved forecasts with observations from system $Y$. We declare that $X$ is not closed, as it is receiving influence (or information) from system $Y$, when data from $Y$ improves forecasts of $X$. Such a reduction of uncertainty perspective of causal inference is not identical to the interventionists' concept of allowing perturbations and experiments  to decide what changes indicate influences.  This data oriented philosophy of causal inference is especially appropriate when (1) the system is a dynamical system of some form producing data streams in time, and (2) a score of influence may be needed.  In particular, contrasting forecasts is the defining concept underlying Granger Causality (G-causality) and it is closely related to the concept of information flow as defined by transfer entropy \cite{TESchreiber2000,TEErik}, which can be proved as a nonlinear version of Granger's otherwise linear (ARMA) test \cite{GCausTERel2009}.  In this spirit we find methods such as Convergent Cross-Mapping method (CCM) \cite{CCMSug2012}, and causation entropy (CSE) \cite{CSESUN2014} to disambiguate direct versus indirect influences, \cite{CSESUN2014,CSESunE2015, bollt2018introduction, runge2019inferring,  lord2016inference,  kim2017causation, almomani2019entropic, sudu2016information}. On the other hand, closely related to information flow are concepts of counter factuals: ``what would happen if.." \cite{nedhall} that are foundational questions for another school leading to the highly successful J. Pearl ``Do-Calculus" built on a specialized variation of Bayesian analysis, \cite{Pearl2001-3}.  These are especially relevant for nondynamical questions (inputs and outputs occur once across populations), such as a typical question of the sort, ``why did I get fat " may be premised on inferring probabilities of removing influences of saturated fats and chocolates. However, with concepts of counter-factual analysis in mind, one may argue that Granger is less descriptive of causation inference, but rather more descriptive of information flow.  In fact, there is a link between the two notions for so-called ``settable" systems under a conditional form of exogeneity \cite{Causationintro1, settableSys1}.

\begin{figure}[hbtp]
\centering
\resizebox{0.5\textwidth}{!}{
\begin{tikzpicture}[->,>=stealth']
 \node[state,text width=4.6cm] (G)
 {\begin{tabular}{l}
  \textbf{\Large Geometry}\\
  \parbox{4.6cm}{\begin{itemize}[leftmargin=*]
      \item[-] \Large Dimensionality
      \item[-] \Large Information flow structure
      \item[-] \Large Level sets
  \end{itemize}}
 \end{tabular}};
 \node[state,
  right of=G,
   node distance=10cm,
  anchor=center,
  text width=4.6cm] (C)
 {
 \begin{tabular}{l}
  \textbf{\Large Causation (Granger)}\\
  \parbox{4.6cm}{\begin{itemize}[leftmargin=*]
      \item[-] \Large Transfer entropy
      \item[-]\Large Geometric Causation
  \end{itemize}}
 \end{tabular}
 };

 \path 
  (C)   edge                node[anchor=left,right, above]{Eq (\ref{geocauC}), (\ref{relTG})} (G);
\end{tikzpicture}
}
\caption{Summary of the paper and relationship of causation and geometry. }
\label{bigpic}
\end{figure}
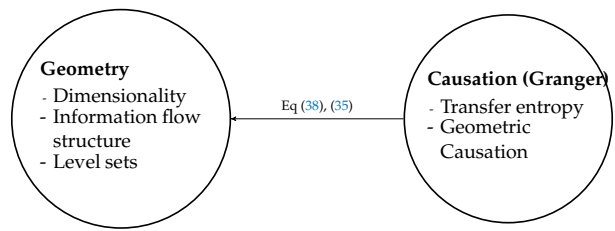

This paper focuses on the information flow perspective, which is causation as it relates to  G-causality.  The role of this paper is to highlight connections between the probabilistic aspects of information flow, such as Granger causality and transfer entropy, to a less often discussed geometric picture that may underlie the information flow.  To this purpose, here we develop both analysis and data driven concepts to serve in bridging what have otherwise been separate philosophies. Figure.~\ref{bigpic} illustrates the two nodes that we tackle here: causal inference and geometry. In the diagram, the equations that are most central in serving to bridge the main concepts are highlighted, and the main role of this paper then could be described as building these bridges. 

When data is derived from a stochastic or deterministic dynamical system, one should also be able to understand the connections between variables in geometric terms. The traditional narrative of information flow is in terms of comparing stochastic processes in probabilistic terms. However, the role of this paper is to offer a unifying description for interpreting  geometric formulations of causation together with traditional statistical or information theoretic interpretations. Thus we will try to provide a bridge between concepts of causality as information flow to the underlying geometry since geometry is perhaps a natural place to describe a dynamical system.

Our work herein comes in two parts.  First, we analyze connections between information flow by transfer entropy to geometric quantities that describe the orientation of underlying functions of a corresponding dynamical system.  In the course of this analysis, we have needed to develop a new ``asymmetric transfer operator" (asymmetric Frobenius-Perron operator) evolving ensemble densities of initial conditions between spaces whose dimensionalities do not match. With this, we proceed to give a new exact formula for transfer entropy, and from there we are able to relate this Kullback-Leibler divergence based measure directly to other more geometrically relevant divergences, specifically total variation divergence and Hellinger divergence, by  Pinsker's inequality.  This leads to a succinct upper bound of the transfer entropy by quantities related to a more geometric description of the underlying dynamical system.  In the second part of this work,  we present numerical interpretations of  transfer entropy $TE_{y\rightarrow x}$ in the setting of a succession of simple dynamical systems, with specifically designed underlying densities, and eventually we include a heart rate versus breathing rate data set.  Then we present a new measure in the spirit of G-causality that is more directly motivated by geometry. This measure, $GeoC_{y\rightarrow x}$, is developed in terms of the classical fractal dimension concept of correlation dimension.

In summary the main theme of this work is to provide connections between probabilistic interpretations and geometric interpretations of causal inference.  The main connections and corresponding sections of this paper are summarized as a dichotomy: Geometry and Causation (information flow structure) as described in Fig.~ (\ref{bigpic}).  Our contribution in this paper is as follows:

\begin{itemize}
    \item In traditional methods, causality is estimated by probabilistic terms. In this study we present analytical and data driven approach to identify causality by geometric methods, and thus also a unifying perspective.
    \item We show that derivative (if it exits) of the underlining function of the time series has a close relationship to the transfer  entropy. (Section \ref{secGeoVsProb})
    \item We provide a new tool called $geoC$ to identify the causality by geometric terms. (Section \ref{secGeoCaus})
    \item Correlation dimension can be used as a measurement for dynamics of a dynamical system. We will show that  this measurement can be used to identify the causality. (Section \ref{secGeoCaus})
\end{itemize}

\part*{\large Part I: Analysis of Connections Between Probabilistic Methods and Geometric Interpretations}
 
%%%%%%%%%%%%%%%%%%%%%%%%%%%%%%%%%%%%%%%%%%
\section{The Problem Setup}\label{secSetup}

 For now, we assume that $x,y$ are real valued scalars, but the multi-variate scenario will be discussed subsequently. We use a shorthand notation, $x:=x_n$, $x':=x_{n+1}$ for any particular time $n$, where the  prime ($'$) notation denotes ``next iterate".  Likewise, let $z=(x,y)$ denote the composite variable, and  its future composite state, $z'$.
Consider the simplest of cases, where there are two coupled dynamical systems written as discrete time maps,
\begin{eqnarray}\label{ds1}
x'&=& f_1(x,y),  \label{eq1} \\
y'&=&f_2(x,y).  \label{eq2}
\end{eqnarray}

The definition of transfer entropy, \cite{TESchreiber2000, TEErik, SyNBollt2012}, measuring the influence of coupling from variables $y$ onto the future of the variables $x$, denoted $x'$ is g.ven by:
\begin{equation}\label{Tyx}
T_{y \rightarrow x}=D_{KL}(p(x'|x)||p(x'|x,y)).
\end{equation}
This hinges on the contrast between two alternative versions of the possible origins of $x'$ and is premised on deciding one of the following two cases: Either
\begin{equation}\label{decide}
    x'=f_1(x),\ \ \mbox{ or }\ \ x'=f_1(x,y),
\end{equation}
is descriptive of the actual function $f_1$.  The definition of $T_{y \rightarrow x}$ is defined to decide this question 
by comparing the deviation from a proposed Markov property,

\begin{equation}\label{markovprop}
    p(x'|x) \stackrel{?}{=} p(x'|x,y).
\end{equation}

The Kullback-Leibler divergence used here contrasts these two possible explanations  of the process generating $x'$.  Since $D_{KL}$ may be written in terms of mutual information, the units are as any entropy, bits per time step.
Notice that we have overloaded the notation writing $p(x'|x)$ and $p(x'|x,y)$.  Our practice will be to rely on the arguments to distinguish functions as otherwise different (likewise distinguishing cases of $f_1(x)$ versus $f_1(x,y)$.   

\begin{figure}[hbtp]
\centering
\includegraphics[width=0.8\textwidth]{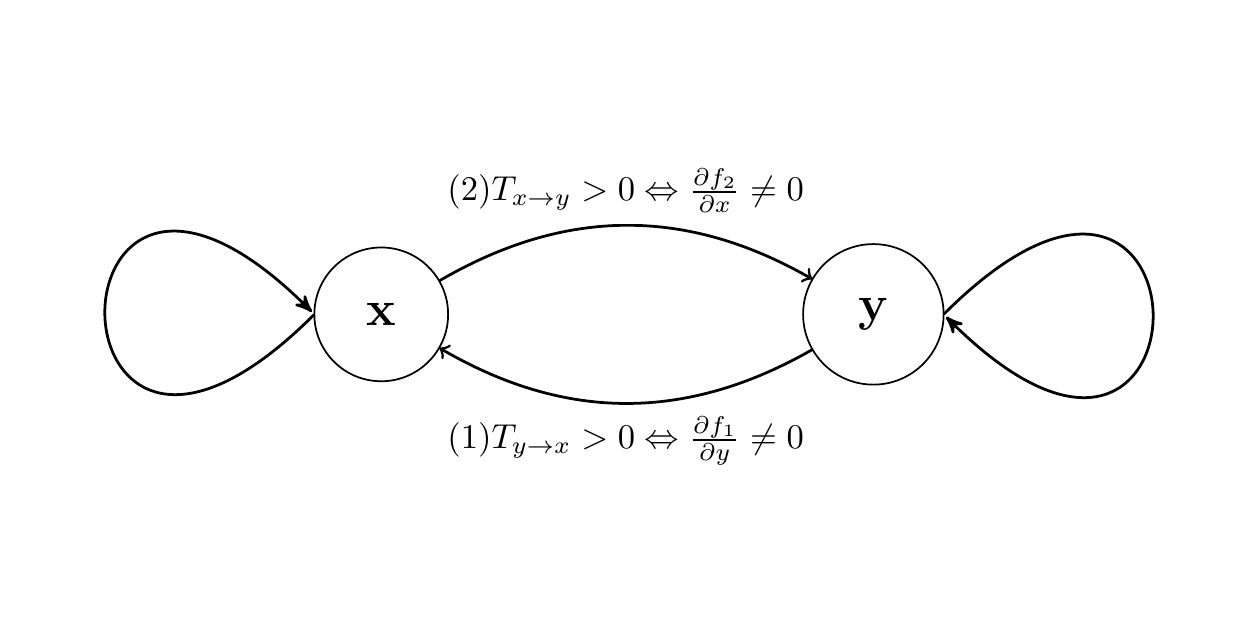}
\caption{A directed graph presentation of the coupling stucture questions corresponding to Eqs.~(\ref{eq1})-(\ref{eq2}).}
\label{fig1}
\end{figure}

Consider that the coupling structure between variables may be caricatured by the directed graph illustrated in Fig.~\ref{fig1}.
In one time step, without loss of generality, we may decide Eq.~(\ref{decide}), the role of $y$ on $x'$, based on $T_{y \rightarrow x}>0$, exclusively in terms of the details of the argument structure of $f_1$. This is separate from the reverse question of $f_2$ as to whether $T_{x \rightarrow y}>0$.  In geometric terms, assuming $f_1\in C^1(\Omega_1)$, it is clear that unless the partial derivative $\frac{\partial f_1}{\partial y}$ is zero everywhere, then the $y$ argument in $f_1(x,y)$ is relevant.  This is not a necessary condition for $T_{y \rightarrow x}>0$ which is a probabilistic statement, and almost everywhere is sufficient.

\subsection{In Geometric Terms}

\begin{figure}[hbtp]
                \centering
             \begin{subfigure}[b]{.49\textwidth}
             \centering
  \includegraphics[width=1\textwidth]{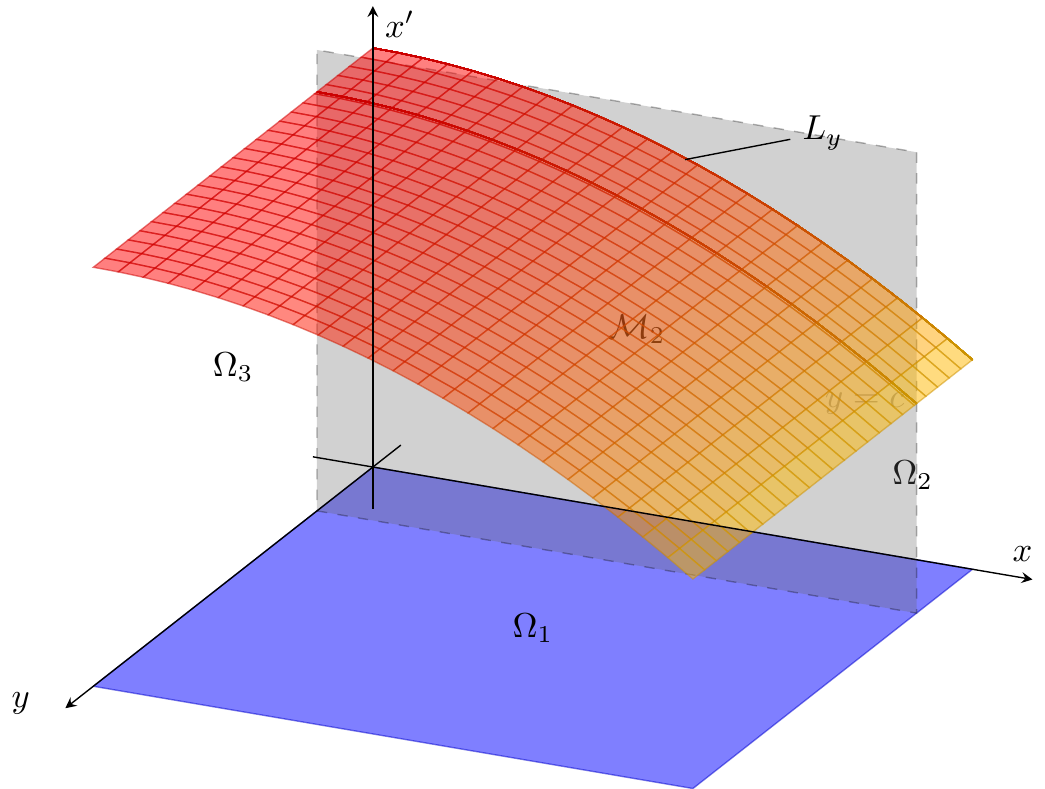}
            \caption{}
            \end{subfigure}
            \begin{subfigure}[b]{0.49\textwidth}
           \centering  \includegraphics[width=1\textwidth]{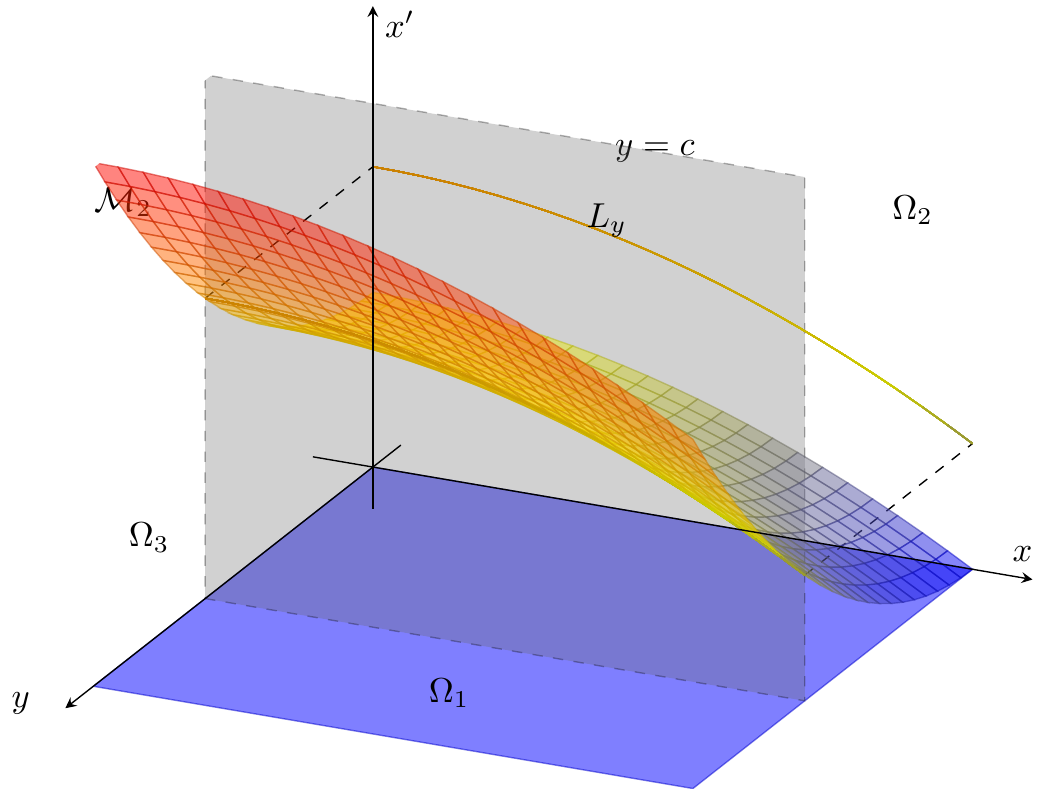}
            \caption{}
            \end{subfigure}
  \caption{$\Omega_2=X\times X'$ manifold and $L_y$ level set for  (a) $x'=f_1(x)=-0.005x^2+100$, (b) $x'=f_1(x,y)=-0.005x^2+0.01y^2+50$. The dimension of the projected set of $(x,x')$ depends on the causality as just described. Compare to Fig.~\ref{fig3}, and Eq.~(\ref{fpimpalance}).} 
\label{fig2}
        \end{figure}

Consider a manifold of points, $(x,y,x')\in X\times Y \times X'$ as the graph over $\Omega_1$, which we label ${\cal M}_2$.  In the following we assume $f_1\in C^1(\Omega_1), \Omega_1\subset X\times Y$.  Our primary assertion here is that the geometric aspects of the  set $(x,y,x')$ projected into $(x,x')$ distinguishes the information flow structure.
Refer to Fig.~\ref{fig2} for notation.
Let the level set for a given fixed $y$ be defined,
\begin{equation}
L_y:=\{(x,x'):x'=f(x,y), y=constant\}\in \Omega_2=X\times X'
\end{equation}

When these level sets are distinct, then the question of the relevance of $y$ to the outcome of $x'$ is clear:
\begin{itemize}
    \item If $\frac{\partial f_1}{\partial y} = 0$ for all $(x,y)\in \Omega_1$, then $L_y=L_{\tilde{y}}$ for all $y,\tilde{y}$.
\end{itemize}
Notice that if the $y$ argument is not relevant as described above,  then $x'=f_1(x)$ better describes the associations, but if we nonetheless insist to write $x'=f_1(x,y)$, then $\frac{\partial f_1}{\partial y} = 0$ for all $(x,y)\in \Omega_1$.  The converse is interesting to state explicitly,
\begin{itemize}
    \item If $L_y\neq L_{\tilde{y}}$ for some $y,\tilde{y}$, then $\frac{\partial f_1}{\partial y} \neq 0$ for some $(x,y)\in \Omega_1$, and then $x'=f_1(x)$ is not a sufficient description of what should really be written $x'=f_1(x,y)$.  We have assumed $f_1\in C^1(\Omega_1)$ throughout.
\end{itemize}

\subsection{In Probabilistic Terms}\label{probterms}

Considering the evolution of $x$ as a stochastic process \cite{LasotaStocP,TEErik}, we may write a probability density function in terms of all those variables that may be relevant, $p(x,y,x')$. 
To contrast the role of the various input variables requires  us to develop a new  singular transfer operator between domains that do not necessarily have the same number of variables. Notice that the definition of transfer entropy  (Eq.~ \ref{Tyx})  seems to rely on the absolute continuity of the joint probability density $p(x,y,x')$. 
However, that  joint distribution of $p(x,y,f(x,y))$
is generally not absolutely continuous, noticing its support is $\{(x,y,f(x,y)): (x,y)\in \Omega_x\times \Omega_y\subseteq \mathbb{R}^2\}$ a measure $0$ subset of $\mathbb{R}^3$. Therefore, the expression $h(f(X,Y)|X,Y)$ is not well defined as a differential entropy and hence there is a problem with transfer entropy.  We expand upon this important detail in the upcoming subsection. To guarantee existence, we interpret these quantities by convolution to smooth the problem. Adding an ``artificial noise" with standard deviation parameter $\epsilon$ allows definition of the conditional entropy at the singular limit $\epsilon$ approaches to zero, and likewise the transfer entropy follows.

The probability density function of the sum of two continuous random variables ($U,Z$) can be obtained by convolution, $P_{U+Z}=P_U* P_Z$.
Random noise ($Z$ with mean $\mathbb{E}(Z)=0$ and  variance $\mathbb{V}(Z)=C\epsilon^2$) added to the original observable variables regularizes,  and we are interested in the singular limit,  $\epsilon\to 0$. We assume that $Z$ is independent of $X,Y$. In experimental data from  practical problems, we argue that some noise, perhaps even if small, is always present. Additionally, noise is assumed to be uniform or normally distributed in practical applications. Therefore, simplicity of the discussion we mostly focused in those two distribution. With this concept, Transfer Entropy can now be calculated by using $h(X'|X,Y)$ and $h(X'|X)$ when
\begin{align}
    X'=f(X,Y)+Z,
\end{align}
where now we assume that $X,Y,Z\in\mathbb{R}$ are independent random variables and we assume that $f:\Omega_x\times \Omega_y \to \mathbb{R}$ is a component-wise monotonic (we will consider the monotonically increasing case for consistent explanations, but one can use monotonically decreasing functions in similar manner) continuous function of $X,Y$ and $\Omega_x,\Omega_y \subseteq \mathbb{R}$. 

\subsubsection{Relative Entropy for a Function of Random Variables}

Calculation of transfer entropy depends on the conditional probability. Hence we will first focus on conditional probability. Since for any particular values $x,y$, the function value $f(x,y)$ is fixed, we conclude that $X'|x,y$ is just a linear function of $Z$. We  see that
\begin{align}
    p_{X'|X,Y}(x'|x,y)=Pr(Z=x'-f(x,y))=p_Z(x'-f(x,y)),
\end{align}
where $p_Z$ is the probability density function of $Z$.

Note that the random variable $X'|x$ is a function of $(Y,Z)$. To write  $U+Z$, let  $U=f(x,Y)$. Therefore convolution of densities of $U$ and $Z$ gives the density function for $p(x'|x)$ (See section \ref{resTrans1} for examples). Notice that a given value of the random variable, say $X=\alpha$, is a parameter in $U$. Therefore, we will denote $U=f(Y;\alpha)$. We will first focus on the probability density function  of $U$,  $p_U(u)$, using the Frobenius-Perron operator,

\begin{align}\label{conprob1}
    %p_{U}(u)= \frac{p_Y(f^{-1}(u;\alpha))}{|f'(f^{-1}(u;\alpha))|}.
    p_{U}(u)=\sum_{y:u=f(y;\alpha)} \frac{p_Y(f(y;\alpha))}{|f'(f(y;\alpha))|}.
\end{align}

In the multivariate setting,  the formula is extended similarly interpreting the derivative as the Jacobian matrix, and the absolute value is interpreted as the absolute value of the determinant. Denote $\mathbf{Y}=(Y_1,\ Y_2,\ \dots,\ Y_n),\  \mathbf{g(\mathbf{Y}};\alpha)=(g_1,\ g_2,\ \dots,\ g_n)$ and  $U=f(\alpha,\mathbf{Y}):=g_1(\mathbf{Y};\alpha)$; and the vector $\mathbf{V}=(V_1,\ V_2,\ \dots,\ V_{n-1})\in \mathbb{R}^{n-1}$ such that $V_i=g_{i+1}(\mathbf{Y}):=Y_{i+1}$ for $i=1,\ 2,\ \dots,\ n-1$. Then the absolute value of the determinate of the Jacobian matrix is given by: $|J_g(\mathbf{y})|=|\frac{\partial g_1(\mathbf{y};\alpha)}{\partial y_1}|$.  As an aside, note that $J$ is  lower triangular with diagonal entries $d_{ii}=1$ for $i>1$.  The  probability density function of $U$ is given by
\begin{align}\label{conprob2}
    p_{U}(u)=\int_S p_{\mathbf{Y}}(g^{-1}(u,\mathbf{v};\alpha))\Big|\frac{\partial g_1}{\partial y_1}(g^{-1}(u,\mathbf{v};\alpha))\Big|^{-1} d\mathbf{v},
\end{align}
where $S$ is the support set of the random variable $\mathbf{V}$.

Since the random variable $X'|x$ can be written as a sum of $U$ and $Z$, we find the probability density function by convolution as follows:
\begin{align}\label{conupz}
 p_{X'|x}(x'|x)=\int p_U(u)p_Z(x'-u)du.   
\end{align}

Now the conditional differential entropy $h(Z|X,Y)$ is in terms of  these probability densities. It is useful that translation does not change the differential entropy, $h_{\epsilon}(f(X,Y)+Z|X,Y)=h(Z|X,Y)$. Also $Z$ is independent from $X,Y$,  $h(Z|X,Y)=h(Z)$.
 Now, we define
\begin{align}\label{lmtDep}
    h(f(X,Y)|X,Y):= \lim_{\epsilon\to 0^+} h_{\epsilon}(f(X,Y)+Z|X,Y)
\end{align}
if this limit exist. 

We consider two scenarios: (1) $Z$ is a uniform random variable or (2) $Z$ is a Gaussian random variable. If it is uniform in the interval $[-\epsilon/2,\epsilon/2],$ then  the differential entropy is $h(Z)=\ln(\epsilon)$.  If specifically, $Z$ is Gaussian with zero mean and $\epsilon$ standard deviation, then  $h(Z)=\frac{1}{2}\ln(2\pi e \epsilon^2)$. Therefore $h_{\epsilon}(f(X,Y)+Z|X,Y) \to -\infty$ as $\epsilon \to 0^+$ in both cases. Therefore, in the $h(f(X,Y)|X,Y))$ is not finite in this definition (Eq.~\ref{lmtDep}) as well. So instead of calculating $X'=f(X,Y)$,  we need to use a noisy version of data $X'=f(X,Y)+Z$.  For that case,
\begin{align}
    h(X'|X,Y)=h(Z)=\begin{cases}
    \ln(\epsilon); & Z\sim U(-\epsilon/2,\epsilon/2)\\
    \frac{1}{2}\ln(2\pi e \epsilon^2); & Z\sim \mathcal{N}(0,\epsilon^2)
    \end{cases};
\end{align}
 where $U(-\epsilon/2,\epsilon/2)$ is the uniform distribution in the interval $[-\epsilon/2,\epsilon/2],$ and $\mathcal{N}(0,\epsilon^2)$ is a Gaussian distribution with zero mean and $\epsilon$ standard deviation.

Now, we focus on $h(X'|X)$. If $X'$ is just a function of $X$, then we can similarly  show that: if $X'=f(X),$ then
\begin{align}
    h(f(X)+Z|X)=h(Z)=\begin{cases}
    \ln(\epsilon); & Z\sim U(-\epsilon/2,\epsilon/2)\\
    \frac{1}{2}\ln(2\pi e \epsilon^2); & Z\sim \mathcal{N}(0,\epsilon^2).
    \end{cases}
\end{align}

Also notice that if $X'=f(X,Y)$ then $h(X'|X)$ will exist, and most of the cases will be finite. But when we calculate $T_{y\to x}$ we need to use the noisy version to avoid the issues in calculating $h(X'|X,Y)$. We will now, consider the interesting case  $X'=f(X,Y)+Z$ and calculate $h(X'|X)$. We require $p_{X'|X}$ and Eq.~(\ref{conupz}) can be used to calculate this probability. Let us denote $I:=\int p_U(u)p_Z(x'-u)du,$ then
\begin{align}
    h_{\epsilon}(X'|X)&=\int\int I\ p_{X}(x)\ln(I)dx'dx\\ \nonumber
    &=\int p_X(x)\int I \ln(I)dx'dx\\
    &=\mathbb{E}_{X}(Q) \nonumber,
\end{align}
where $Q=\int I \ln(I)dx'$. Notice that if $Q$ does not depend on $x,$ then $h(X'|X)=Q\int p_X dx=Q$ because $\int p_X dx=1$(since $p_x$ is a probability density function). Therefore, we can calculate $h_{\epsilon}(X'|X)$ by four steps. First we calculate the density function for $U=f(x,Y)$ (by using Eq.~(\ref{conprob1}) or (\ref{conprob2}) ). Then, we calculate $I=p_{X'|X}$ by using Eq.~(\ref{conupz}).  Next, we calculate the value of $Q,$ and finally we calculate the value of $h_{\epsilon}(X'|X)$.

Thus  the transfer entropy from $y$ to $x$ follows in terms of comparing conditional entropies,
\begin{align}\label{TxyconEntr}
    T_{y\to x}=h(X'|X)-h(X'|X,Y).
\end{align}
This quantity is not well defined when $X'=f(X,Y)$, and therefore we considered the $X'=f(X,Y)+Z$ case. This interpretation of transfer entropy depends on the parameter $\epsilon$, as we  define,
\begin{align}
    T_{y\to x}:=\lim_{\epsilon\to 0^+}T_{y\to x}(\epsilon) =\lim_{\epsilon\to 0^+} h_{\epsilon}(X'|X)-h_{\epsilon}(X'|X,Y)
\end{align}
if this limit exist.

Note that,
\begin{align}
    T_{y\to x}=\begin{cases}\lim_{\epsilon\to 0^+} h(Z)-h(Z)=0;\ & X'=f(X)\\
    \infty;\ & X'=f(X,Y)\ne f(X).
    \end{cases}
\end{align}
Thus we see that a finite quantity is ensured by the noise term. We can easily find an upper bound for the transfer entropy when $X'=f(X,Y)+Z$ is a random variable with finite support (with all the other assumptions mentioned earlier) and  suppose $Z\sim U(-\epsilon/2,\epsilon/2)$. First, notice that the uniform distribution maximizes entropy amongst all distributions of continuous random variables with finite support. If $f$ is component-wise monotonically increasing continuous function then the support of $X'|x$ is $[f(x,y_{min})-\epsilon/2, f(x,y_{min})+\epsilon/2]$ for all $x\in \Omega_x$. Here $y_{min}$ and $y_{max}$ are minimum and maximum values of $Y$. Then it follows that
\begin{align}
    h_{\epsilon}(X'|X)\le ln(|f(x_{max},y_{max})-f(x_{max},y_{min})+\epsilon|),
\end{align}
where $x_{max}$ is the maximum $x$ value. We see that an interesting upper bound for transfer entropy follows:
\begin{align}
    T_{y\to x}(\epsilon)\le \ln\Big(\Big|\frac{f(x_{max},y_{max})-f(x_{max},y_{min})}{\epsilon}+1\Big|\Big).
\end{align}
%%%%%End of New(Sep 8, 2019)%%%%%%%%%%%%%%%%%%%

\subsection{Relating Transfer Entropy to  a Geometric Bound}{\label{secGeoVsProb}}
Noting that transfer entropy and other variations of the G-causality concept are expressed in terms of conditional probabilities, we recall that,
\begin{equation}
\rho(x'|x,y){\rho(x,y)}={\rho(x,y,x')}.
\end{equation}
Again we continue to overload the notation on the functions $\rho$, the details of the arguments distinguishing to which of these functions we refer. 

Now consider the change of random variable formulas that map between probability density functions by smooth transformations.  In the case that $x'=f_1(x)$ (in the special case that $f_1$ is one-one) then
\begin{equation}
\rho(x')=\frac{\rho(x)}{|\frac{d{f_1}}{{dx}}(x)|}=\frac{\rho(f_1^{-1}(x'))}{|\frac{d{f_1}}{{dx}}(f_1^{-1}(x'))|}.
\end{equation}
In the more general case, not assuming one-one-ness, we get the usual Frobenius-Perron operator,
\begin{equation}\label{FP1}
\rho(x')=\sum_{x:x'=f_1(x)}\rho(x,x')=\sum_{x:x'=f_1(x)}\frac{\rho(x)}{|\frac{d{f_1}}{{dx}}(x)|},
\end{equation}
in terms of a summation over all pre-images of $x'$.  Notice also that the middle form is written as a marginalization across $x$ of all those $x$ that lead to $x'$. This Frobenius-Perron operator,  as usual, maps densities of ensembles of initial conditions under the action of the map $f_1$.

Comparing to the expression
\begin{equation}
    \rho(x,x')=\rho(x'|x) \rho(x),
\end{equation}
we assert the interpretation that
\begin{equation}\label{interp1}
    \rho(x'|x):=\frac{1}{|\frac{d{f_1}}{{dx}}(x)|}\delta(x'-f_1(x)),
\end{equation}
where $\delta$ is the Dirac delta function.
In the language of Bayesian uncertainty propagation, $p(x'|x)$ describes the likelihood function, if interpreting the future state $x'$ as data, and the past state $x$ as parameters, in a standard Bayes description, $ p(\mbox{data}|\mbox{parameter}) \times p(\mbox{parameter})$.  As usual for any likelihood function, while it is a probability distribution over the data argument, it may not necessarily be so with respect to the parameter argument. 

Now consider the case where $x'$ is indeed nontrivially a function with respect to not just $x$, but also with respect to $y$.  Then we require the following asymmetric space transfer operator, which we name here an asymmetric  Frobenius-Perron operator for smooth transformations between spaces of dissimilar dimensionality:
\begin{Theorem}[Asymmetric Space Transfer Operator ]\label{ThmASTO}
 If $x'=f_1(x,y)$, for $f_1:\Omega_1 \rightarrow \Upsilon$, given bounded open domain $(x,y)\in \Omega_1\subset {\mathbb R}^{2d}$, and range $x'\in \Upsilon\subset {\mathbb R}^{d}$, and $f_1\in C^1(\Omega_1)$, and the Jacobian matrices, $\frac{\partial{f_1}}{{\partial x}}(x,y)$, and $\frac{\partial{f_1}}{{\partial y}}(x,y)$ are not both rank deficient at the same time, then taking the initial density
 $\rho(x,y)\in L^1(\Omega_1)$,  the following serves as a transfer operator mapping asymmetrically defined densities $P:L^1(\Omega_1)\rightarrow L^1(\Upsilon)$
\begin{equation}
\rho(x')=\sum_{(x,y):x'=f_1(x,y)}\rho(x,y,x')=\sum_{(x,y):x'=f_1(x,y)}\frac{\rho(x,y)}{|\frac{\partial{f_1}}{{\partial x}}(x,y)|+|\frac{\partial{f_1}}{{\partial y}}(x,y)|}.
\end{equation}
\end{Theorem}

The proof of this is in Appendix \ref{dis}. Note also that by similar argumentation, one can formulate the asymmetric Frobenius-Perron type operator between sets of disimilar dimensionality in an integral form.

\begin{figure}[hbtp]
\centering
\includegraphics[width=0.6\textwidth]{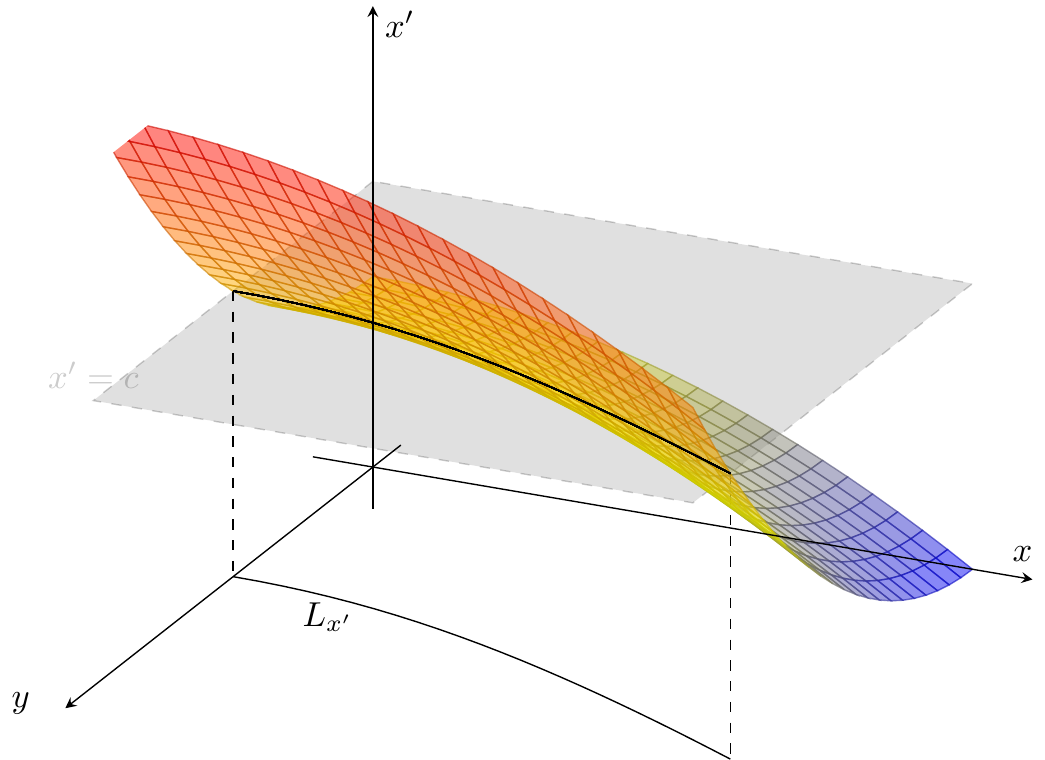}
\caption{The asymmetric transfer operator, Eq.~(\ref{fpimpalance}),  is written in terms of intefration over the level set, $L_{x'}$ of $x'=f_1(x,y)$ associated with a fixed value $x'$, Eq.~(\ref{levelsetLxp}).}
\label{fig3}
\end{figure}

\begin{Corollary}[Asymmetric Transfer Operator, Kernel Integral Form]\label{CorATOKI}
Under the same hypothesis as Theorem \ref{ThmASTO}, we may alternatively write the integral kernel form of the expression,
\begin{eqnarray}\label{fpimpalance}
P:L^2({\mathbb R}^2)&\rightarrow L^2({\mathbb R}) \\
\rho(x,y)&\mapsto & \rho'(x')= P[\rho](x,y)]\nonumber \\ &=&  \nonumber \\
&=&\int_{L_{x'}}\rho(x,y,x')dx dy=\int_{L_{x'}}\rho(x'|x,y)\rho(x,y) dx dy \nonumber \\ &=& \int_{L_{x'}} \frac{1}{|\frac{\partial{f_1}}{{\partial x}}(x,y)|+|\frac{\partial{f_1}}{{\partial y}}(x,y)|} \rho(x,y) dx dy .
\end{eqnarray}
This is in terms of a line integration along the level set, ${L_{x'}}$.  See Fig.~\ref{fig3}.
\begin{equation}\label{levelsetLxp}
    L_{x'}=\{(x,y)\in \Omega_1: f(x,y)=x' \mbox{ a chosen constant.}\}
\end{equation}
\end{Corollary}

In Fig.~\ref{fig3}, we have shown a typical scenario where a level set is a curve (or it may well be a union of disjoint curves), whereas in a typical FP-operator between sets of the same dimensionality generally the integration is between pre-images that are usually either singletons, or unions of such points, $\rho'(x')=\int \delta(s-f(x)) \rho(s) ds=\sum_{x:f(x)=x'} \frac{\rho(x)}{|Df(x)|}$.

Contrasting standard and the asymmetric forms of transfer operators as described above, in the next section we will compute and bound estimates for the transfer entropy.  However, it should already be apparent that, if $\frac{\partial f_1}{\partial y}=0$ in probability with respect to $\rho(x,y)$, then $T_{y\rightarrow x}=0$.

\medskip

{\bf Contrast to other statistical divergences reveals geometric relevance: }
Information flow is quite naturally defined by the KL-divergence,  in that it comes in the units of entropy, e.g. bits per second. However, the well known Pinsker's inequality \cite{pinsker1960information}  allows us to more easily relate the transfer entropy to a quantity that  has a geometric relevance  using the total variation, even if this is only by an inequality estimate.

Recall Pinsker's inequality \cite{pinsker1960information} relates random variables with probability distributions $p$ and $q$ over the same support to the total variation and the KL-divergence as follows,
\begin{equation}
   0\leq  \frac{1}{2} TV(P,Q)\leq \sqrt{D_{KL}(P||Q)},
\end{equation}
written as probability measures $P$, $Q$.  The total variation distance between probability measures is a maximal absolute difference of possible events, 
\begin{equation}
TV(P,Q)=\sup_A |P(A)-Q(A)|,
\end{equation}
but it is well known to be related to 1/2 of the L1-distance 
in the case of a common dominating measure, $p(x)d\mu=dP$, $q(x) d\mu=dQ$.  In this work, we only need  absolute continuity with respect to Lebesgue measure, $p(x)=dP(x), q(x)=dQ(x)$; then,
\begin{equation}
TV(P,Q)  =\frac{1}{2}\int |p(x)-q(x)|dx=\frac{1}{2} \|p-q\|_{L^1},
\end{equation}
here with respect to Lebesgue measure.
Also, we write $D_{KL}(P||Q)=\int p(x)\log \frac{p(x)}{q(x)} dx$,  therefore, 
\begin{equation}
    \frac{1}{2}\|p-q\|_{L^1}^2\leq \int p(x)\log \frac{p(x)}{q(x)} dx.
\end{equation}

Thus, with the Pinsker inequality, we can bound the transfer entropy from below by inserting the definition Eq.~(\ref{Tyx}) into the above:
\begin{equation}
     0\leq \frac{1}{2} \|p(x'|x,y)-p(x'|x)\|_{L^1}^2\leq T_{y\rightarrow x}.
\end{equation}
The assumption that the two distributions correspond to a common dominating measure requires that we interpret $p(x'|x)$ as a distribution averaged across the same $\rho(x,y)$ as $p(x'|x,y)$.  (Recall by definition \cite{boucheron2013concentration} that $\lambda$ is a common dominating measure of $P$ and $Q$ if $p(x)=dP/d\lambda$ and $q(x)=dQ/d\lambda$ describe corresponding densities).     For the sake of simplification we interpret transfer entropy relative to a uniform initial density, $\rho(x,y)$ for both entropies of Eq.~(\ref{TxyconEntr}). With this assumption we interpret 
\begin{equation}\label{relTG}
    0\leq \frac{1}{2} \| \frac{1}{|\frac{\partial{f_1}}{{\partial x}}(x,y)|+|\frac{\partial{f_1}}{{\partial y}}(x,y)|}
    -
    \frac{1}{|\frac{d{f_1}}{{dx}}(x)|}
    \|_{{L^1}(\Omega_1,\rho(x,y))}^2\leq T_{y\rightarrow x}.
\end{equation}
In the special case that there is very little information flow, we would expect that $|\frac{\partial f_1}{\partial y}|<b<<1$, and $b<<|\frac{\partial f_1}{\partial x}|$, a.e. $x,y$; then a power series expansion in small $b$ gives
\begin{equation}\label{approxsmallb}
\frac{1}{2} \| \frac{1}{|\frac{\partial{f_1}}{{\partial x}}(x,y)|+|\frac{\partial{f_1}}{{\partial y}}(x,y)|}
    -
    \frac{1}{|\frac{d{f_1}}{{dx}}(x)|}
    \|_{L^1(\Omega_1,\rho(x,y))}^2  \approx \frac{Vol(\Omega_1)}{2} 
\frac{<|\frac{\partial f_1}{\partial y}|>^2}{<|\frac{\partial f_1}{\partial x}|>^4},
\end{equation}
which serves approximately as the TV-lower bound for transfer entropy where have used the notation $<\cdot>$ to denote an average across the domain.  Notice that therefore, $\delta( p(x'|x,y),p(x'|x) ) \downarrow$ as $|\frac{\partial f_1}{\partial y}|\downarrow$. While Pinsker's inequality cannot guarantee that therefore $ T_{y\rightarrow x}\downarrow$, since TV is only an upper bound, it is clearly suggestive. In summary, comparing inequality Eq.~(\ref{relTG}) to the approximation (\ref{approxsmallb}) suggests that for 
$|\frac{\partial f_1}{\partial y}|<<b<<|\frac{\partial f_1}{\partial x}|$, for $b>0$, for a.e. $x,y$, then $ T_{y\rightarrow x}\downarrow$ as $b\downarrow $.

Now, we change to a more computational direction of this story of interpreting information flow in geometric terms.  With the strong connection described in the following section we bring to the problem of information flow between geometric concepts  to information flow concepts, such as entropy, it is natural to turn to studying the dimensionality of the outcome spaces, as we will now develop.

%%%%%%%%G-casation%%%%%%%

\part*{ \large Part II: Numerics and Examples of Geometric Interpretations}

Now we will explore numerical estimation aspects of transfer entropy for causation inference in relationship to geometry as described theoretically in the previous section, and we will compare this numerical approach to geometric aspects.

\section{Geometry of Information Flow }\label{secGeoCaus}
As theory suggests, see above sections, there is a strong relationship between the information flow (causality as measured by transfer entropy) and the geometry, encoded for example in the estimates  leading to Eq.~(\ref{approxsmallb}). The effective dimensionality  of the underlying manifold as projected into the outcome space is a key factor to identify the causal inference between chosen variables. Indeed any question of causality is in fact observer dependent. To this point, suppose $x'$ only depends on $x,\ y$ and $x'=f(x,y)$ where $f\in C^1(\Omega_1)$. We noticed that (Section \ref{secSetup}) $T_{y\to x} =0 \iff \frac{\partial f}{\partial y}= 0,\  \forall (x,y)\in \Omega_1$.  Now notice that $\frac{\partial f}{\partial y}= 0,\  \forall (x,y)\in \Omega_1 \iff x'=f(x,y)=f(x).$ Therefore, in the case that $\Omega_1$ is  two dimensional, then $(x,x')$ would be a one dimensional manifold if and only if $\frac{\partial f}{\partial y}= 0,\  \forall (x,y)\in \Omega_1$.  See Fig.~\ref{fig2}.  With these assumptions, \[T_{y\to x}=0 \iff (x,x') \text{ data lie on a } 1-D \text{ manifold}.\]
Likewise, for more general dimensionality of the initial $\Omega_1$, the story of the information flow between variables is in part a story of how the image manifold is projected.
Therefore, our discussion will focus on estimating the dimensionality in order to identify the nature of the underlying manifold. Then, we will focus on identifying causality by estimating the dimension of the manifold, or even more generally of the resulting set if it is not a manifold but perhaps even a fractal. Finally, this naturally leads us to introduce  a new geometric measure for characterizing the causation, which we will identify as $Geo_{y\rightarrow x}$. 
\subsection{Relating the Information Flow as Geometric Orientation of Data.}
For a given time series $x:=x_n\in \mathbb{R}^{d_1},y:=y_n\in \mathbb{R}^{d_2}$, consider the $x':=x_{n+1}$ and {\it contrast} the dimensionalities of $(x,y,x')$ versus $(x,x')$,  to identify that $x'=f(x)$ or $x'=f(x,y)$. 
Thus, in mimicking the premise of Granger causality, or likewise of Transfer entropy, contrasting these two versions of the explanations of $x'$, in terms of either $(x,y)$ or $x$ we decide the causal inference, but this time, by using only the geometric interpretation. First we recall how fractal dimensionality evolves under transformations, \cite{DimAnaYorke1991}. 
\begin{Theorem}[ \cite{DimAnaYorke1991}]\label{dimTheorem} Let $A$ be a bounded Borel subset of $\mathbb{R}^{d_1}$. Consider the function $F:A\to \mathbb{R}^{d_1}\times \mathbb{R}^{d_1}$ such that $F(x)=(x, x')$ for some $x'\in \mathbb{R}^{d_1}$. The correlation dimension $D_2(F(A))\le d_1$, if and only if there exists a function $f:A \to \mathbb{R}^{d_1}$ such that $x'=f(x)$ with $f\in C^1(A)$. 
\end{Theorem}
The idea of the arguments in the complete proof found in Sauer et.~ al., \cite{DimAnaYorke1991}, are as follows. Let $A$ be bounded Borel subset of $\mathbb{R}^{d_1}$ and $f:A \to \mathbb{R}^{d_1}$ with $f\in C^1(A)$. Then $D_2(f(A))=D_2(A)$ where $D_2$ is the correlation dimension \cite{DimPreYorke1997}.  Note  that $D_2(A)\le d_1$. Therefore $D_2(F(A))=D_2(A)\le d_1$, with $F:A\to \mathbb{R}^{d_1}\times\mathbb{R}^{d_1}$ if and only if  $F(x)=(x,\ f(x))$.

Now, we can describe this dimensional statement in terms of our information flow causality discussion, to develop an alternative measure of inference between variables.  Let  $(x,x')\in \Omega_2\subset{\mathbb{R}^{2d_1}}$ and $(x,y,x')\in\Omega_3\subset{\mathbb{R}^{2d_1+d_2}}$. We assert that there is a causal inference from $y$ to $x$, if $dim(\Omega_2)>d_1$ and $d_1<dim(\Omega_3)\le d_1+d_2$, (Theorem 1). In this paper we focus on time series $x_n\in \mathbb{R}$ which might also depend on time series $y_n\in \mathbb{R}$ and we will consider the geometric causation from $y$ to $x$, for $(x,y)\in A\times B =\Omega_1\subset{\mathbb{R}^2}$. We will denote geometric causation by $GeoC_{y\to x}$ and assume that $A,\ B$ are Borel subsets of $\mathbb{R}$. Correlation dimension is used to estimate the dimensionality.
  First, we identify the causality using the dimensionality of on  $(x,x')$ and $(x,y,x')$. Say, for example that  $(x,x')\in \Omega_2\subset{\mathbb{R}^2}$ and $(x,y,x')\in \Omega_3\subset{\mathbb{R}^3}$, then clearly we would enumerate a correlation dimension causal inference from $y$ to $x$, if $dim(\Omega_2)>1$ and $1<dim(\Omega_3)\le 2$, (Theorem 1).

\subsection{Measure Causality by Correlation Dimension}
As we have been discussing. the information flow of a dynamical system can be described geometrically by studying the sets (perhaps they are manifolds) $X\times X'$ and $X\times Y\times X'$.  As we noticed in the last section, comparing the dimension of these sets can be interpreted as descriptive of information flow. Whether dimensionality be estimated from data or by a convenient fractal measure such as the correlation dimension ($D_2(.)$), there is an interpretation of information flow when contrasting $X\times X'$ versus $X\times Y\times X'$, in a spirit reminiscent of what is done with transfer entropy.  However, these details are geometrically more to the point. 

Here, we define $GeoC_{y\to x}$  (geometric information flow) by $GeoC(.|.)$ as conditional correlation dimension. 
\begin{Definition}[Conditional  Correlation Dimensional Geometric Information Flow] Let $\mathcal{M}$ be the manifold of data set $(X_1,\ X_2,\ \dots ,\ X_n,\ X' )$ and let $\Omega_1$ be the data set $(X_1,\ X_2,\ \dots ,\ X_n)$. Suppose that the $\mathcal{M},\ \Omega_1$ are bounded Borel sets. The quantity  \begin{align}\label{geocau2}
    GeoC(X'|X_1,\ \dots ,\ X_{n}):=D_2(\mathcal{M})-D_2(\Omega_1)
\end{align}
is defined as ``Conditional Correlation Dimensional Geometric Information Flow". Here, $D_2(.)$ is the usual correlation dimension of the given set, \cite{GRASSBERGER1983189,PhysRevLett.50.346,GRASSBERGER1983227}.  
\end{Definition}
\begin{Definition}[Correlation Dimensional Geometric Information Flow]
Let $x:=x_n, y=y_n\in \mathbb{R}$ be two time series. The correlation dimensional geometric information flow from $y$ to $x$ as measured by the correlation dimension  and denoted by $GeoC_{y\to x}$ is given by  \begin{align}\label{geocauC} \
    GeoC_{y\to x}:= GeoC(X'|X)-GeoC(X'|X,Y).
\end{align}
\end{Definition}  

A key observation is to notice that, if $X'$ is a function of $(X_1,\ X_2,\ \dots ,\ X_n)$ then $D_2(\mathcal{M})=D_2(\Omega_1)$ otherwise $D_2(\mathcal{M})>D_2(\Omega_1)$ (Theorem 1). If $X$ is not influenced by $y$, then $GeoC(X'|X)=0,$ $GeoC(X'|X,Y)=0$ and therefore $GeoC_{y\to x}=0$. Also, notice that $GeoC_{y\to x}\le D_2(X)$, where $X=\{x_n|n=1,2,\dots\}$. For example if $x_n\in \mathbb{R}$ then $GeoC_{y\to x}\le 1$. Since we assume that influence of any time series $z_n\ne x_n, y_n$ to  $x_n$ is relatively small , we can conclude that $GeoC_{y\to x}\ge 0$, and if $x'=f(x,y)$ then $GeoC(X'|X,Y)=0$. Additionally the dimension ($GeoC(X'|X)$) in the $(X,X')$ data scores how much additional (other than $X$) information is needed to describe $X'$ variable. Similarly, the dimension $GeoC(X'|X,Y)$ in the $(X,Y,X')$ data describes how much additional (other than $X,Y$) information is needed to define $X'$. However, when the number of data points $N \to \infty$, the value $GeoC_{y\to x}$ is nonegative (equal to the dimension of $X$ data). Thus, theoretically $GeoC$ identifies  a causality in the geometric sense we have been describing.

\section{Results and discussion}

Now, we present specific examples to contrast the transfer entropy with our proposed geometric measure to further highlight the role of geometry in such questions. Table.~\ref{table:sumery} provides summary of our numerical results. We use synthetic examples with known underlining dynamics to understand the accuracy of our model. Calculating transfer entropy has theoretical and numerical issues for those chosen examples while our geometric approach accurately identifies the causation. We use the correlation dimension of the data because data might be fractals. Using H\'{e}non map example, We demonstrate that fractal data will not affect our calculations. Furthermore, we use a real-world application that has a positive transfer entropy to explain our data-driven geometric method. Details of these examples can be found in the following subsections.

\begin{table}[H]
\caption{Summery of the results. Here we experiment our new approach by synthetics and real world application data.}
\centering 

\begin{tabular}{p{2in}| p{1.75in}|p{1.25in}}
\hline\hline 
Data  & Transfer Entropy(Sec.~ \ref{resTrans1}) & Geometric Approach \\ [0.5ex]
\hline 
 Synthetic: f(x,y)=$aX+b Y+C$, $a,b,c\in \mathbb{R}$ &  Theoretical issues can be noticed. Numerical estimation have boundedness issues when $b<<1$.& Successfully identify the causation for all the cases ($100\%$).  \\ \hline 
Synthetic: f(x,y)=$ag_1(X)+b g_2(Y)+C$, $a,b,c\in \mathbb{R}$& Theoretical issues can be noticed. Numerical estimation have boundedness issues when $b<<1$.& Successfully identify the causation for all the cases ($100\%$).\\ \hline 
H\'{e}non map: use data set invariant under the map. & special case of $aX^2+bY+C$ with $a=-1.4,\ b=c=1$. Estimated transfer entropy is positive. &  Successfully identify the causation.   \\ \hline 
Application: heart rate vs breathing rate & Positive transfer entropy.& Identify positive causation. It also provide more details about the data.  \\
[1ex] 
\hline \hline 
\end{tabular}
\label{table:sumery} 
\end{table}

\subsection{Transfer Entropy}{\label{resTrans1}}
In this section we will focus on analytical results and numerical estimators for conditional entropy and transfer entropy for specific examples. As we discussed in previous sections starting with \ref{probterms}, computing the  transfer entropy for $X'=f(X,Y)$ has technical difficulties due to the singularity of the quantity $h(X'|X,Y)$. First, we will  consider the calculation of $h(X'|X)$ for $X'=f(X,Y)$, and then we will discuss the calculation for noisy data. In the following examples we assumed that $X,Y$ are random variables such that $X,Y\overset{iid}{\sim} U([1,2])$. A summary of the calculations for a few   examples are listed in the Table \ref{table:hxpx}.
\begin{table}[H]
\caption{Conditional entropy $h(X'|X)$ for $X'=f(X,Y)$, for specific parametric examples listed, under the assumption that $X,Y\overset{iid}{\sim} U([1,2])$.}
\centering 
\begin{tabular}{c| c} 
\hline\hline 
$f(X,Y)$  & $h(X'|X)$ \\ [0.5ex]
\hline 
 $g(X)+b Y$ & $\ln(b)$   \\
$g(X)+b Y^2$  & $\ln(8b)-5/2$ \\
$g(X)+b\ \ln(Y)$ & $\ln(\frac{b\ e}{4})$ \\
[1ex] 
\hline 
\end{tabular}
\label{table:hxpx} 
\end{table}
\begin{figure}[htb]
    \centering
    \begin{subfigure}[b]{\textwidth}
        \centering
        \includegraphics[width=0.49\linewidth]{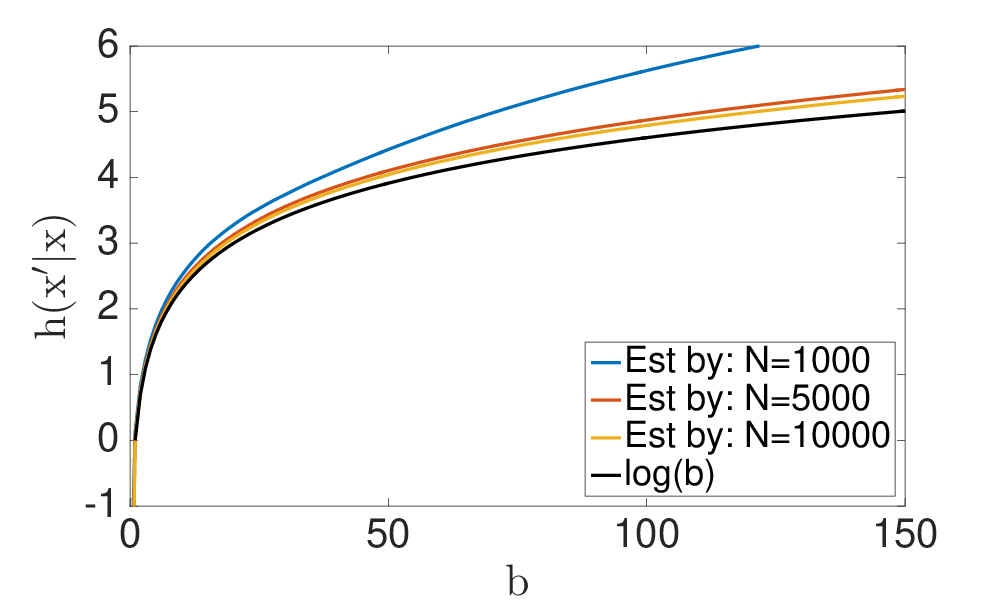}%
        %\hfill
        \includegraphics[width=0.49\linewidth]{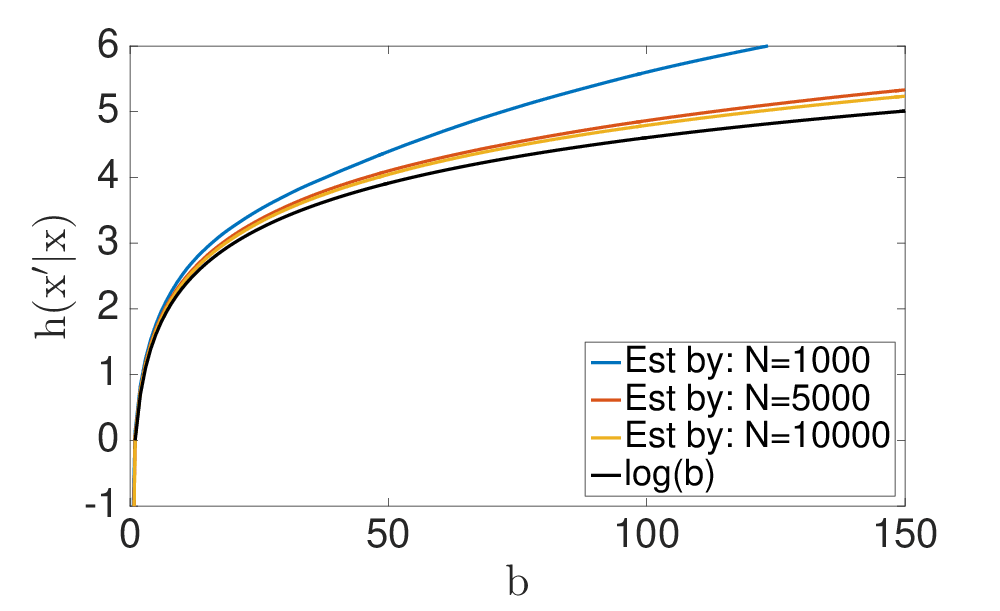}
        \caption{Examples for $X'=g(X)+bY$. Left figure shows result for $g(X)=X$ and right shows result for $g(X)=X^2$. }
    \end{subfigure}
    \vskip\baselineskip
    \begin{subfigure}[b]{\textwidth}
        \centering
        \includegraphics[width=0.49\linewidth]{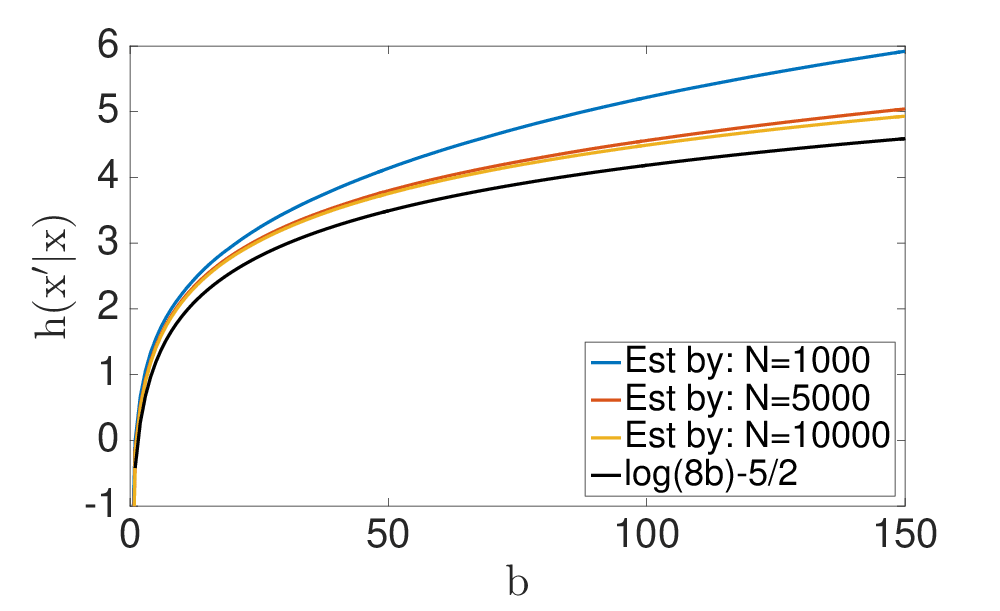}%
        %\hfill
        \includegraphics[width=0.49\linewidth]{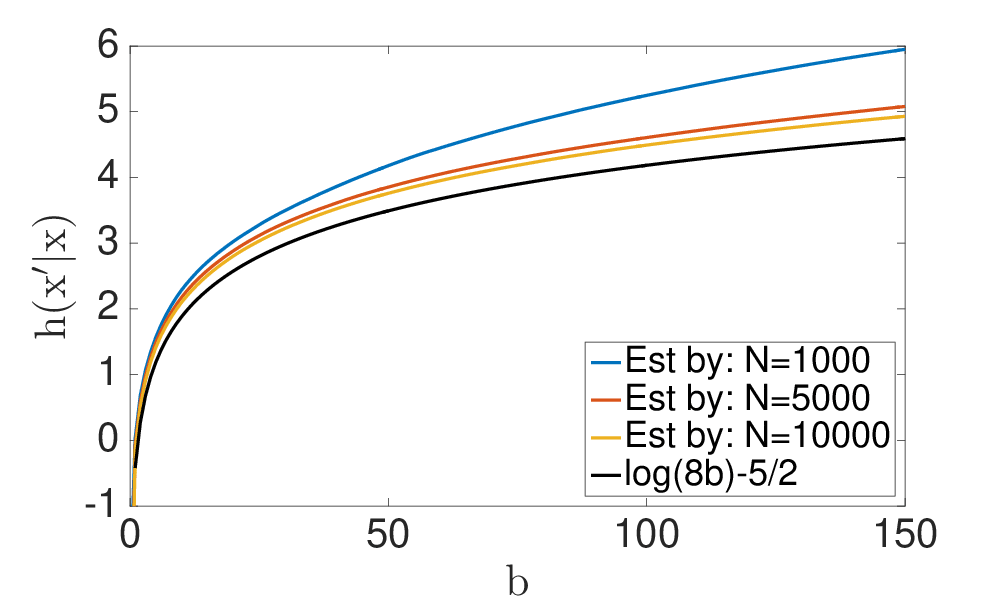}
        \caption{Examples for $X'=g(X)+bY^2$. Left figure shows result for $g(X)=X$ and right shows result for $g(X)=e^x$.}
    \end{subfigure}
    \caption{Conditional entropy $h(X'|X)$. Note that these numerical estimate for the conditional entropy by the KSG method, \cite{PhysRevE.69.066138}, converge(as $N\to \infty$) to the analytic solutions (see Table \ref{table:hxpx}) }
\end{figure}
We will discuss the transfer entropy with noisy data because to make $h(X'|X,Y)$ well defined, requires absolute continuity of the probability density function $p(x,y,x')$. Consider for example the problem  form $X'=g(X)+bY+C$ where $X,Y$ are uniformly distributed independent random variables over the interval $[1,2]$ (the same analysis can be extend to any finite interval) with $b$ being a constant, and $g$ a function of random variable $X$. We will also consider $C$ to be a random variable which is  distributed uniformly on  $[-\epsilon/2,\epsilon/2]$.
Note that it follows that $h(X'|X,Y)=\ln \epsilon$. To calculate the $h(X'|X)$, we need to find the conditional probability $p(X'|x)$ and observe that $X'|x=U+C$ where $U=g(x)+bY$. Therefore, 
\begin{align}
   p_U(u)= \begin{cases}\frac{1}{b} &; g_1(x)+b\le X'\le g_1(x)+2b\\ 0 &; otherwise. \end{cases}
\end{align}
 and 
\begin{align}
   p_{X'|X}(X'|x)= \begin{cases}\frac{x'+\epsilon/2-g(x)}{b\epsilon} &; g(x)-\epsilon/2\le X'\le g(x)+\epsilon/2\\ 
   \frac{1}{b} &; g(x)+\epsilon/2\le X'\le b+g(x)-\epsilon/2\\
   \frac{-x'+\epsilon/2+g(x)+b}{b\epsilon} &; b+g(x)-\epsilon/2\le X'\le b+g(x)+\epsilon/2\\
   0 &; otherwise \end{cases}.
\end{align}
By the definition of transfer entropy we can show that
\begin{align}
    h(X'|X)=\ln b+\frac{\epsilon}{2b}
\end{align}
and hence transfer entropy of this data is given by
\begin{align}\label{trngxpy}
    T_{y\to x}(\epsilon;b)=\begin{cases}
    \ln \frac{b}{\epsilon}+ \frac{\epsilon}{2b}; & b\ne 0\\
    0; &b=0.
    \end{cases}
\end{align}
Therefore,  when $b=0$, the transfer entropy $T_{y\to x}=\ln\epsilon-\ln\epsilon=0$. Also notice that $T_{y\to x}(\epsilon;b)\rightarrow \infty$ as $\epsilon\rightarrow 0$. Therefore convergence of the numerical estimates is  slow when $\epsilon>0$ is small (See Fig.~ \ref{epsForT}). %Therefore numerical solutions for transfer entropy is a clear estimator for the $X'=f(X,Y)$ case. 

\begin{figure}[htb]
    \centering
    \begin{subfigure}[b]{0.49\textwidth}
        \centering
      \includegraphics[width=\linewidth]{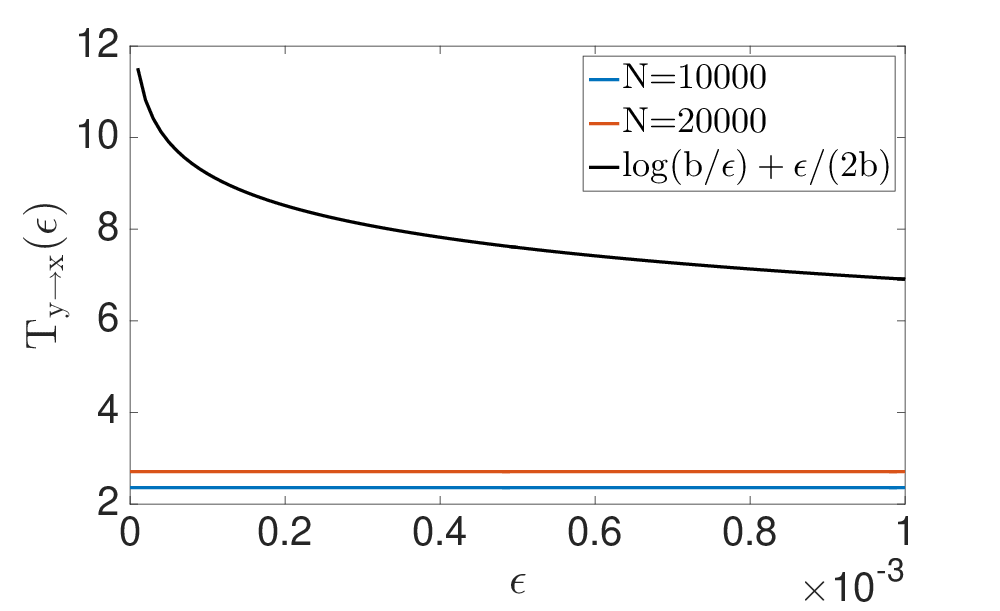}%
        \caption{$b=1$.}
\end{subfigure}

\vspace{1em}
\begin{subfigure}[b]{0.49\textwidth}
        \includegraphics[width=\linewidth]{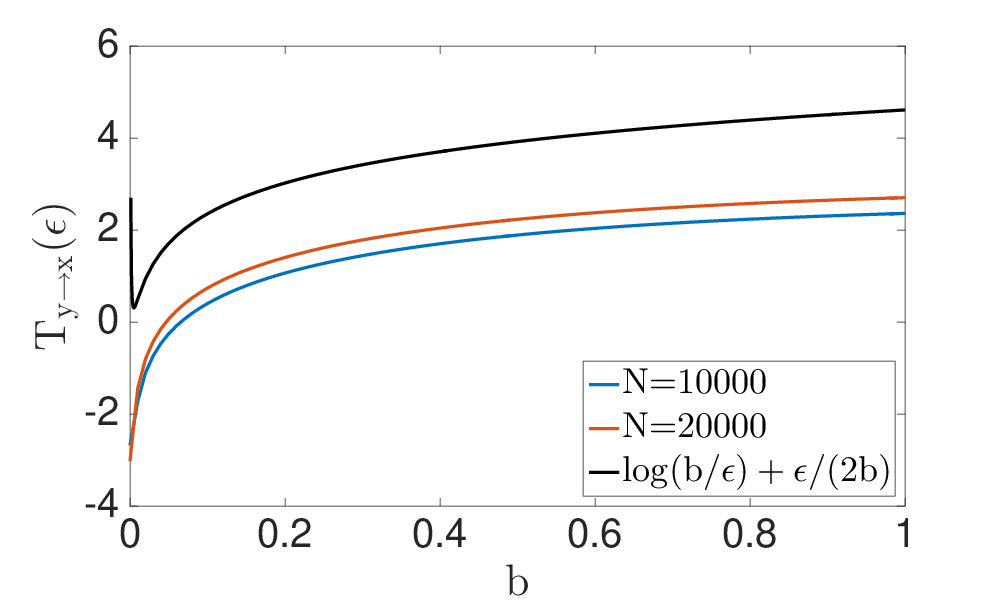}
        \caption{$\epsilon=0.01$}
\end{subfigure}
\begin{subfigure}[b]{0.49\textwidth}
        \includegraphics[width=\linewidth]{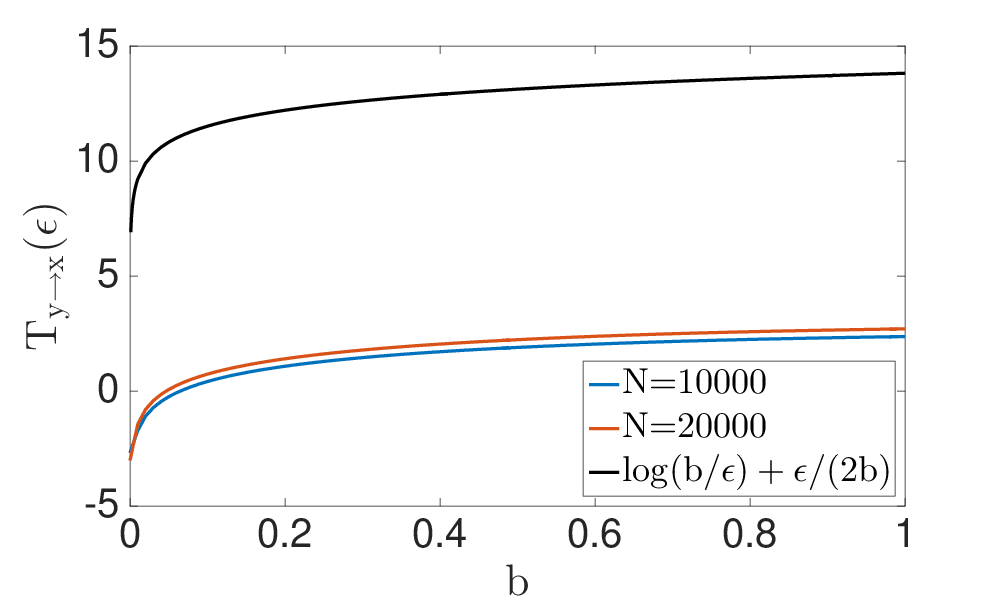}
        \caption{$\epsilon=10^{-6}$}
    \end{subfigure}
    \caption{Numerical results and analytical results for transfer entropy $T_{y\to x}(\epsilon;b)$ to the problem $X'=X+bY+\epsilon$ . Transfer entropy vs $\epsilon$ shows in (a) for fixed $b$ value. (b) and (c) figures shows the behavior of the transfer entropy for $b$ values with fixed $\epsilon$ values. Noticed that convergence of numerical solution is  slow when epsilon is small. }\label{epsForT}
\end{figure}

\subsection{Geometric Information Flow}
Now we  focus on quanitfying the  %diffusion dimensional 
geometric information flow %also 
by comparing dimensionalities of the  outcomes spaces. We will contrast %these 
this to the transfer entropy computations for a few examples of the form $X'=g(X)+bY+C$.

To illustrate the idea of geometric information flow, let us first consider a simple example, $x'=ax+by+c$. If $b=0$, we have $x'=f(x)$ and when $b\ne 0$ we have $x'=f(x,y)$ case. Therefore, dimensionality of the data set $(x', x)$  will change with parameter $b$. When the number of data points $N\to \infty$ and $b\ne0$, then $GeoC_{y\to x} \to 1$. Generally this measure of causality depends on the value of $b$, but also the initial density of initial conditions. 

\begin{figure}[hbtp]
        	\centering
             \begin{subfigure}[b]{.49\textwidth}
             \centering 
  \includegraphics[width=1\textwidth]{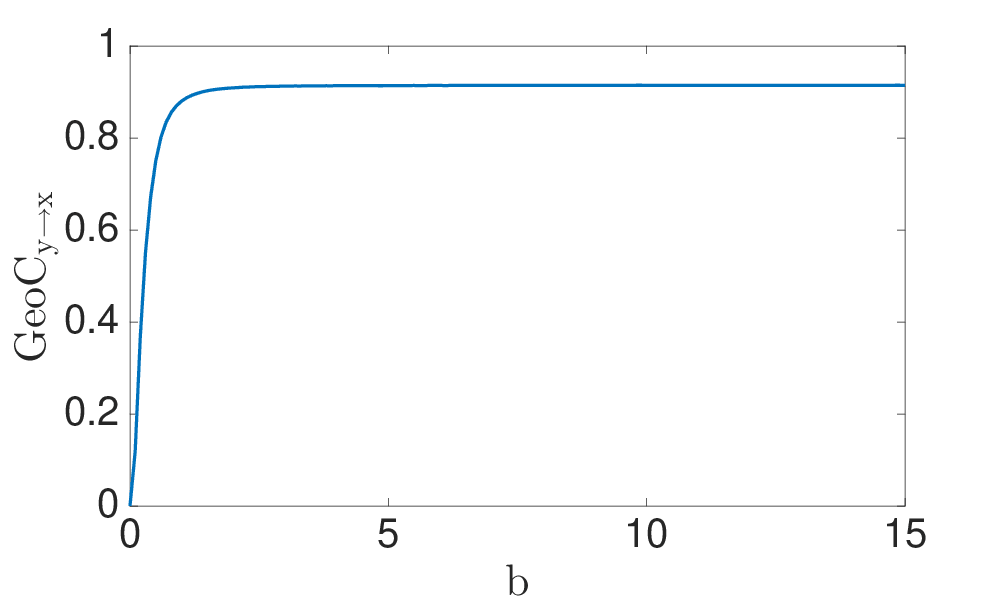}
            \caption{$GeoC_{y\to x}$}
            \end{subfigure}
            \begin{subfigure}[b]{0.49\textwidth}
           \centering  \includegraphics[width=1\textwidth]{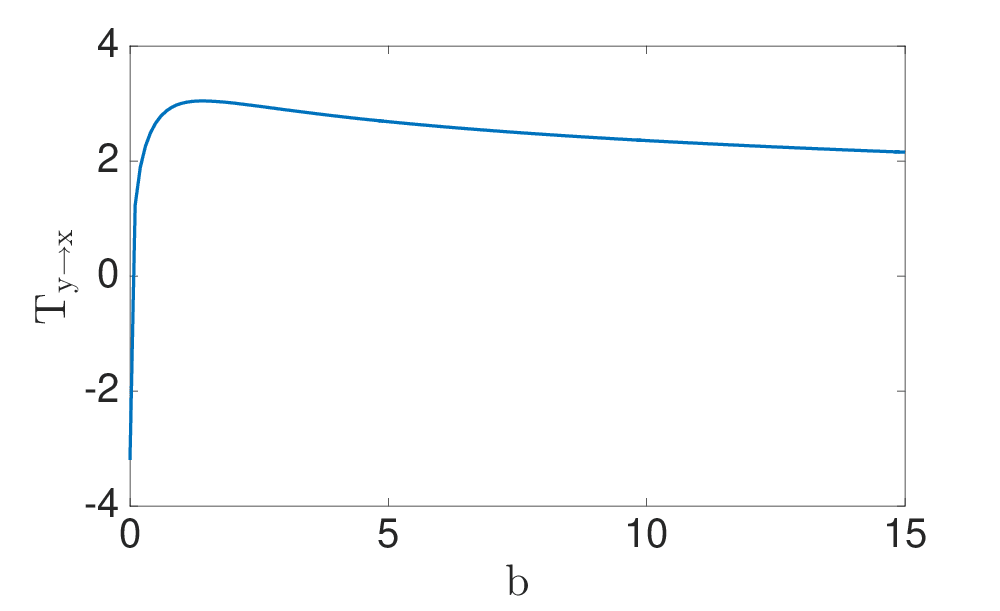}
            \caption{$T_{y\to x}$(Numerical results)}
            \end{subfigure}

            \caption{Geometric information flow vs Transfer entropy for $X'=bY$ data.   
            }\label{GeovsTEby}
        \end{figure}

In this example we contrast theoretical solutions with the numerically estimated solutions, Fig.~\ref{GeovsTEby}.   Theoretically we expect $T_{y\to x}=\begin{cases}0 &; b=0\\ \infty &; b\ne 0\end{cases}$ as $N\to \infty$. Also the transfer entropy for noisy data can be calculated by Eq.~(\ref{trngxpy}).
        
\begin{figure}[hbtp]
        	\centering
             \begin{subfigure}[b]{.49\textwidth}
             \centering 
  \includegraphics[width=1\textwidth,height=45mm,keepaspectratio]{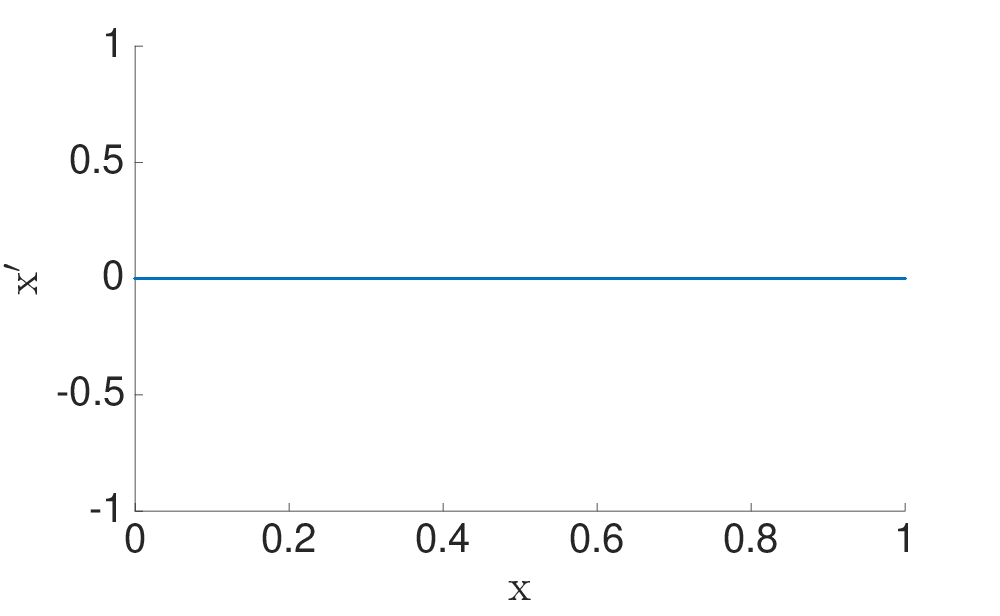}
            \caption{}
            \end{subfigure}
            \begin{subfigure}[b]{0.49\textwidth}
           \centering  \includegraphics[width=1\textwidth,height=45mm,keepaspectratio]{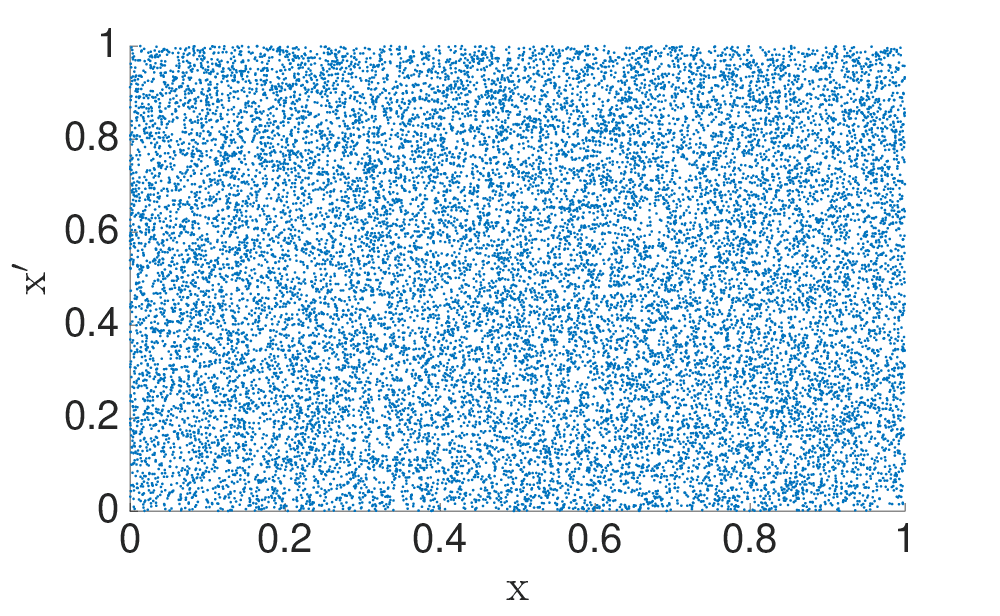}
            \caption{}
            \end{subfigure}

            \caption{ Manifold of the data $(x',x)$ with $x'=by$ and $y$ is uniformly distributed in the interval $[0,1]$.  Noticed that when (a) $b=0$ we have a 1-D manifold, (b) $b\ne 0$ we have 2-D manifold, in the  $(x',x)$ plane. 
            }\label{intro2}
       
        \end{figure}

\subsection{Synthetic data: \texorpdfstring{$X'=aX+bY$ with $a\ne 0$}{Lg}}
The role of the initial density of points in the domain plays an important role in how the specific information flow values are computed depending on the measure used.  To illustrate this point, consider the example of a unit square, $[0,1]^2$, that is uniformly sampled, and mapped by,
\begin{equation}\label{anzatz2}
    X'=aX+bY, \mbox{ with }a\ne 0.
\end{equation} 
    
\begin{figure}[hbtp]
        	\centering
            \begin{subfigure}[b]{.45\textwidth}
             \centering 
  \includegraphics[width=\textwidth]{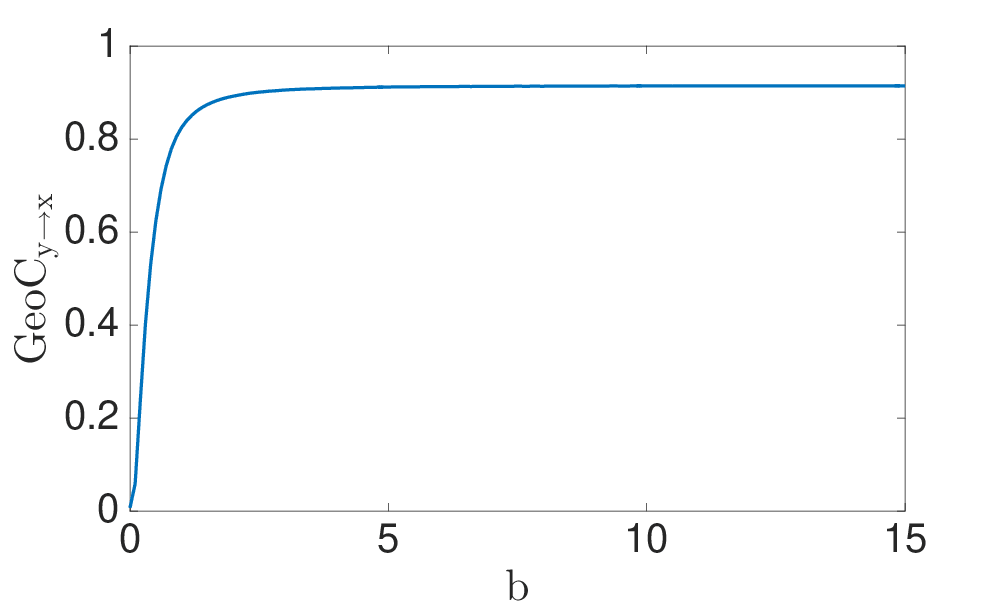}
            \caption{ $GeoC_{y\to x}$ }
            \end{subfigure}
\begin{subfigure}[b]{.45\textwidth}
             \centering 
  \includegraphics[width=\textwidth]{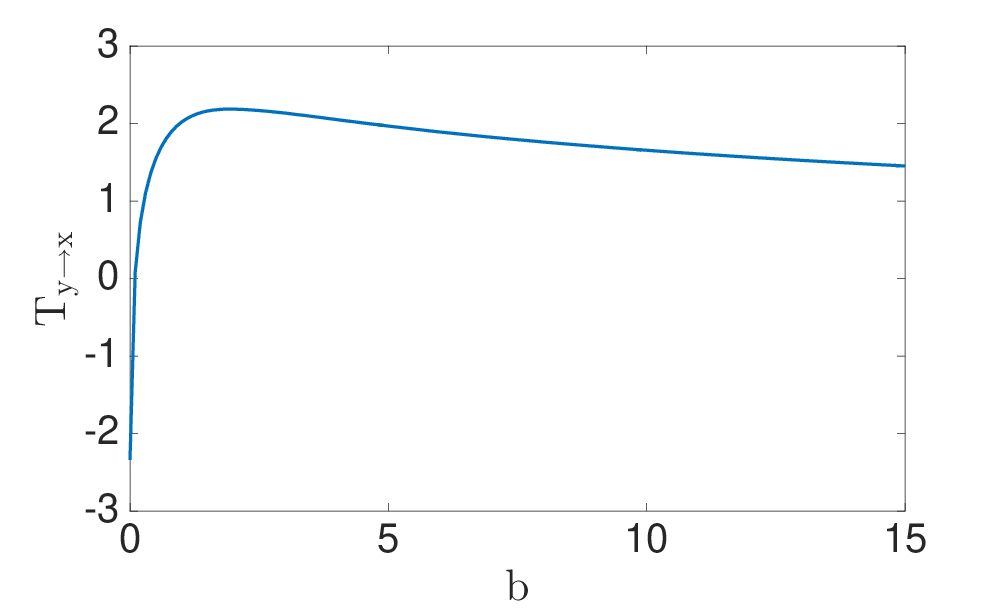}
            \caption{$T_{y\to x}$  }
            \end{subfigure}
\caption{Figure (a) shows the geometric information flow and (b) represent the Transfer entropy for $x'=x+by$ data. Figures shows the changes with parameter $b$. We can noticed the transfer entropy has similar behavior to the geometric information flow of the data.
}\label{GeoVsTrs}
        \end{figure}

This fits our basic premise that $(x,y,x')$ data embeds in a $2$-D manifold, by ansatz of Eqs.~(\ref{eq1}), (\ref{anzatz2}), assuming for this example that each of $x, y$ and $x'$ are scalar. As the number of data point grows,  $N\to \infty$, we can see that $GeoC_{y\to x}=\begin{cases}0 &; b=0\\ 1 &; b\ne 0\end{cases}$, because $ (X,X')$ data is on $2$-D manifold iff $b\ne0$, (numerical estimation can be seen in Fig.~\ref{GeoVsTrs}(b) ). On the other hand, the conditional entropy $h(X'|X,Y)$ is not defined, becoming unbounded when  defined by noisy data. Thus, it follows that transfer entropy shares this same property. In other words, boundedness of transfer entropy depends highly on the $X'|X,Y$  conditional data structure; while instead, our  geometric information flow measure highly depends on $X'|X$ conditional data structure.  Figure.~ \ref{GeoVsTrs}(c) demonstrates this observation with estimated transfer entropy and analytically computed values for noisy data.  The slow convergence can be observed, Eq.~\ref{trngxpy}, Fig.~\ref{epsForT}.
    
\subsection{Synthetic data: Non-linear cases}
  
\begin{figure}[hbtp]
        	\centering
             \begin{subfigure}[b]{.49\textwidth}
             \centering 
  \includegraphics[width=\textwidth]{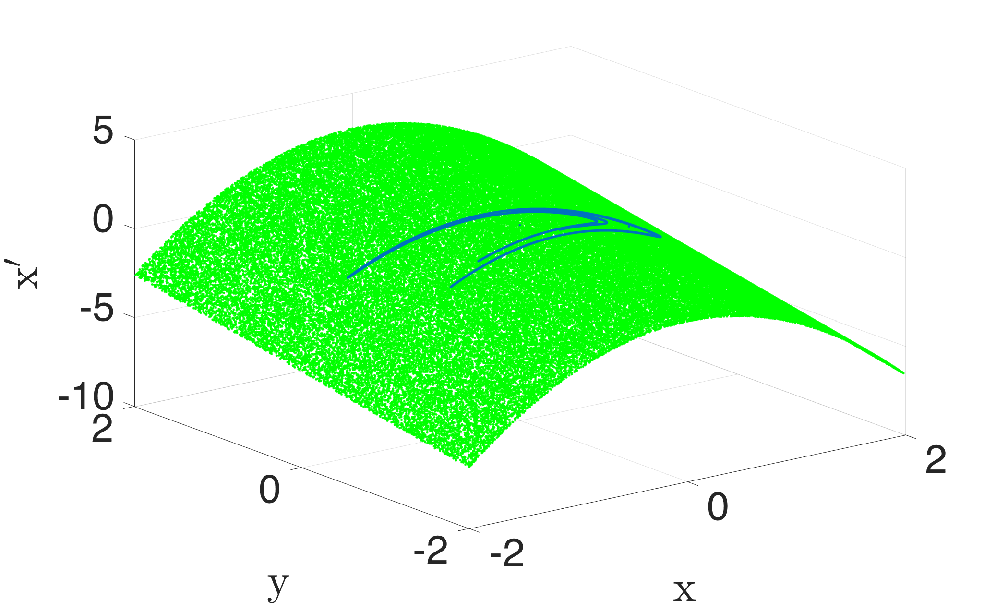}
            \caption{$(x,y,x')$ data for H\'{e}non Map.}
            \end{subfigure}
            
        \vspace{2em}    \begin{subfigure}[b]{0.45\linewidth}
           \centering  \includegraphics[width=\linewidth]{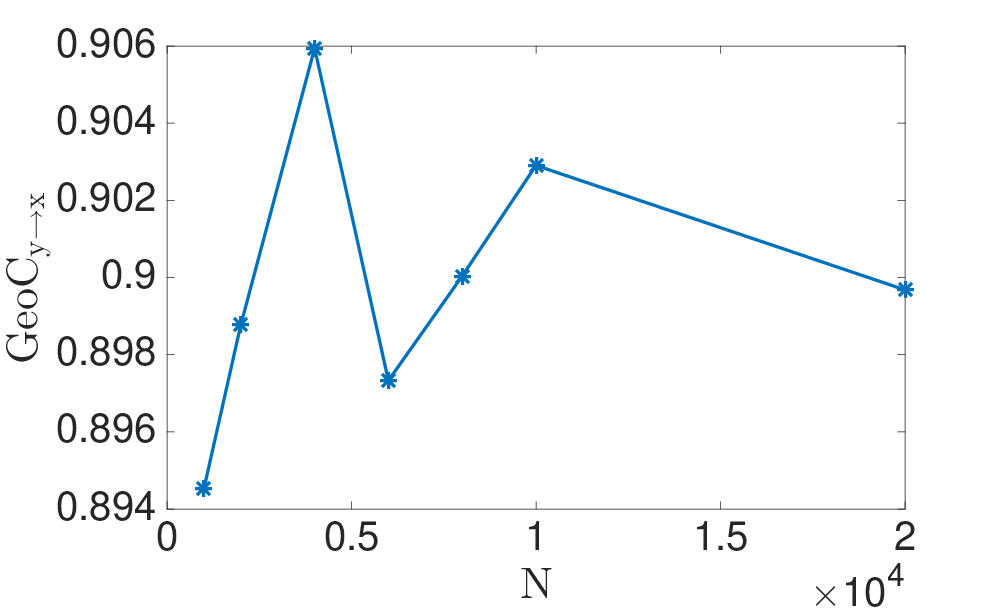}
            \caption{$(x,y)\sim U([-1.5,1.5]^2)$ }
            \end{subfigure}
          \begin{subfigure}[b]{.45\textwidth}
             \centering 
 \includegraphics[width=1\textwidth]{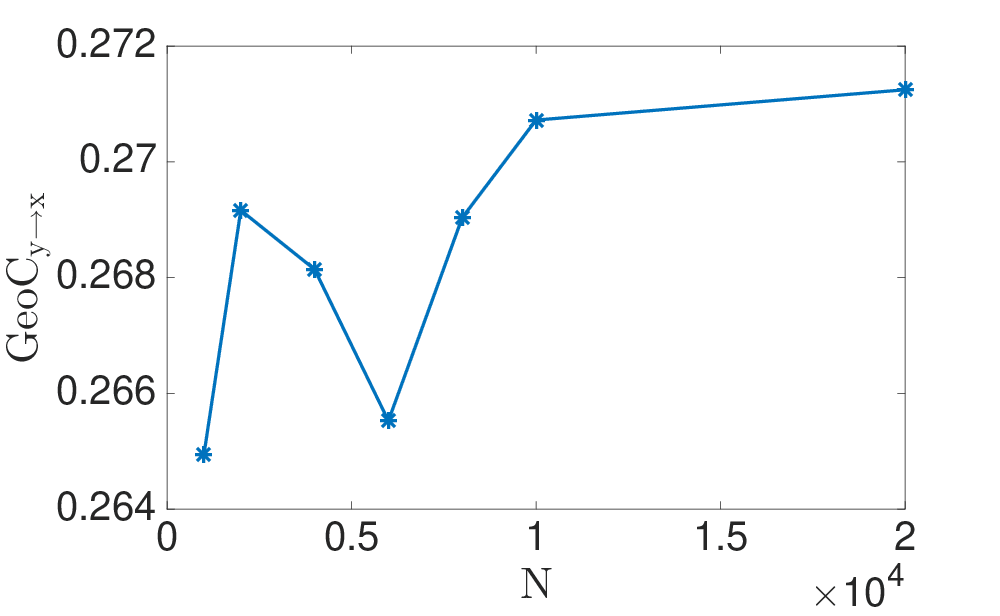}
            \caption{$(x,y)$ is in invariant set of H\'{e}non map}
            \end{subfigure}
            \caption{Consider the H\'{e}non map, Eq.~(\ref{Henoneq}), within the domain  $[-1.5,1.5]^2$ and the invariant set of H\'{e}non map. a) The uniform distribution case (green) as well as the natural invariant measure of the attractor (blue) are shown regarding the $(x,y,x')$ data for both cases.
            b) when $(x,y)\in [-1.5,1.5]^2$ notice that $GeoC_{y\to x}=0.9$ and c) if $(x,y)$ in invariant set of H\'{e}non map, then $GeoC_{y\to x}=0.2712$}\label{phen}
        \end{figure}

Now consider the H\'{e}non map,
\begin{align}\label{Henoneq}
    x'&=1-1.4x^2+y\\ \nonumber
    y'&=x
\end{align}
as a special case of a general quadratic relationship, $x'=ax+by^2+c$, for discussing how $x'$ may depend on $(x,y)\in \Omega_1$.  Again we do not worry here if $y'$ may or may not  depend on $x$ and or $y$ when deciding dependencies for $x'$. We will discuss  two cases, depending on how the $(x,y)\in \Omega_1$ data is distributed. For the first case, assume  $(x,y)$ is uniformly distributed in the square,  $[-1.5,1.5]^2$. The second and dynamically more realistic case will assume  that $(x,y)$ lies on the invariant set (the strange attractor) of the H\'{e}non map. The geometric information flow is shown for both cases, in Fig.~\ref{phen}. We numerically estimate the   transfer entropy for both cases  which gives $T_{y\to x}=2.4116$ and $0.7942$ respectively. (But recall that the first case for transfer entropy might not be finite analytically, and there is slow numerical estimation.),  see Table \ref{table:nonlin}.

\begin{table}[H]
\caption{H\'{e}non Map Results.  Contrasting geometric information flow versus transfer entropy in two diferent cases, 1st relative to uniform distribution of initial conditions (reset each time) and 2nd relative to the natural invariant measure (more realistic).}
\centering 
\begin{tabular}{c c c} 
\hline\hline 
Domain &$GeoC$ & $T_{y\to x}$  \\ [0.5ex]
\hline 
 $[-1.5,1.5]^2$&0.90 & 2.4116  \\ 
Invariant Set  &0.2712& 0.7942  \\
[1ex] 
\hline 
\end{tabular}
\label{table:nonlin} 
\end{table}

\subsection{Application Data }

\begin{figure}[hbtp]
        	\centering
         \begin{subfigure}[b]{0.45\textwidth}
             \centering 
  \includegraphics[width=\linewidth]{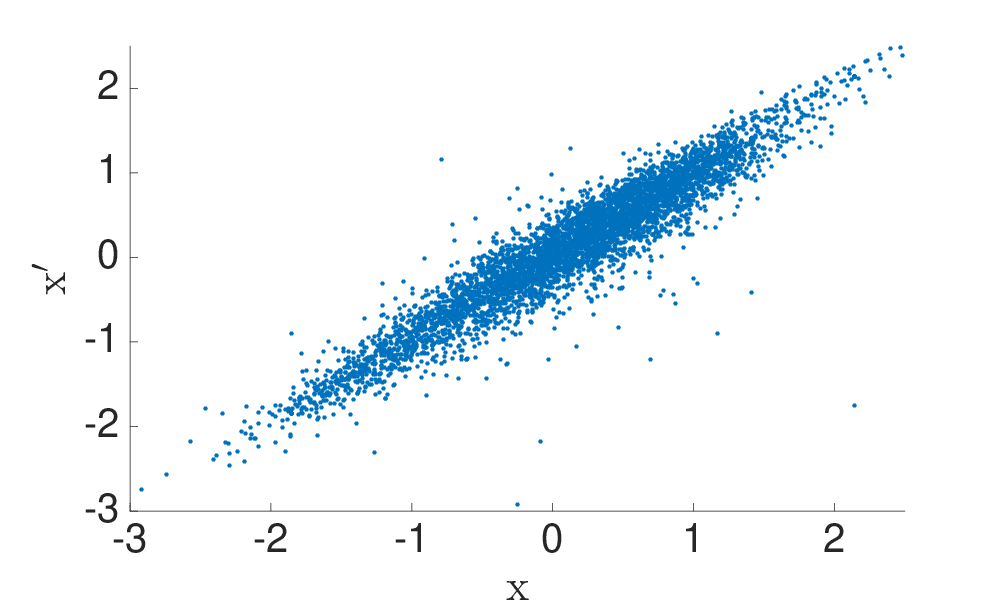}
            \caption{}
            \end{subfigure}
          \begin{subfigure}[b]{0.45\linewidth}
           \centering  \includegraphics[width=\linewidth]{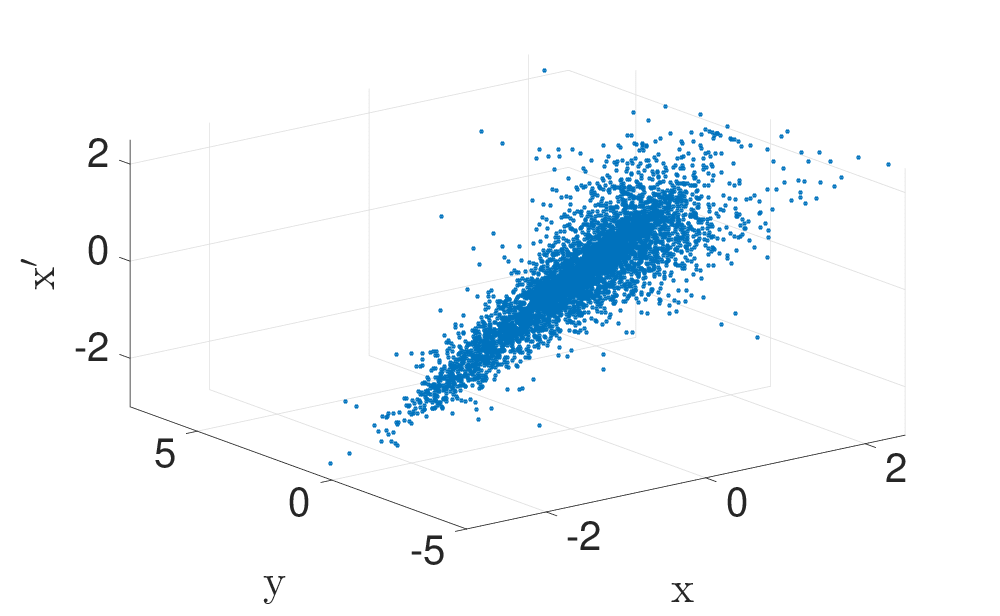}
            \caption{}
            \end{subfigure}
            \vskip\baselineskip
          \begin{subfigure}[b]{0.45\textwidth}
             \centering 
  \includegraphics[width=\linewidth]{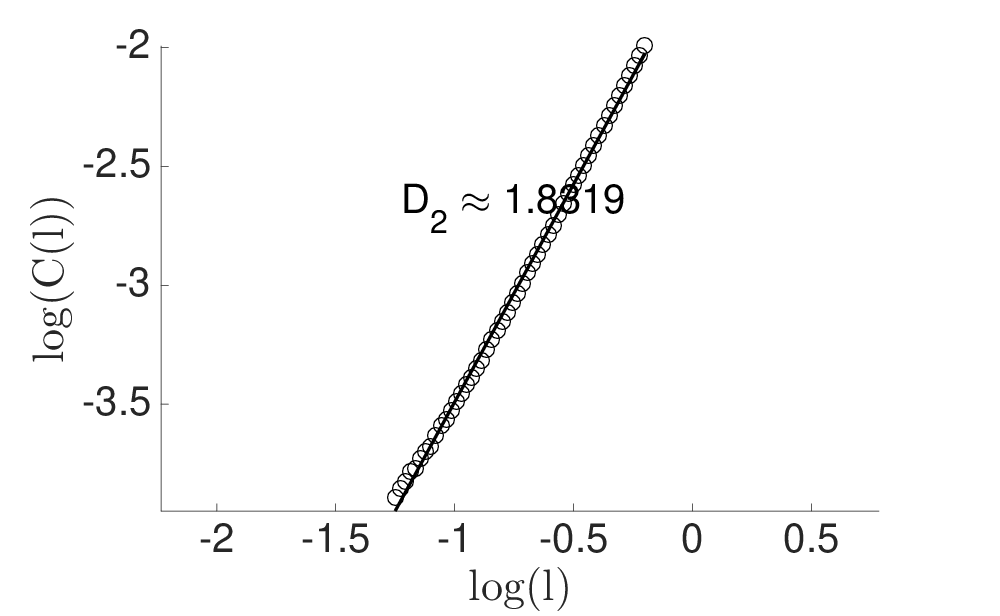}
            \caption{}
            \end{subfigure}
\begin{subfigure}[b]{0.45\textwidth}
             \centering 
  \includegraphics[width=\linewidth]{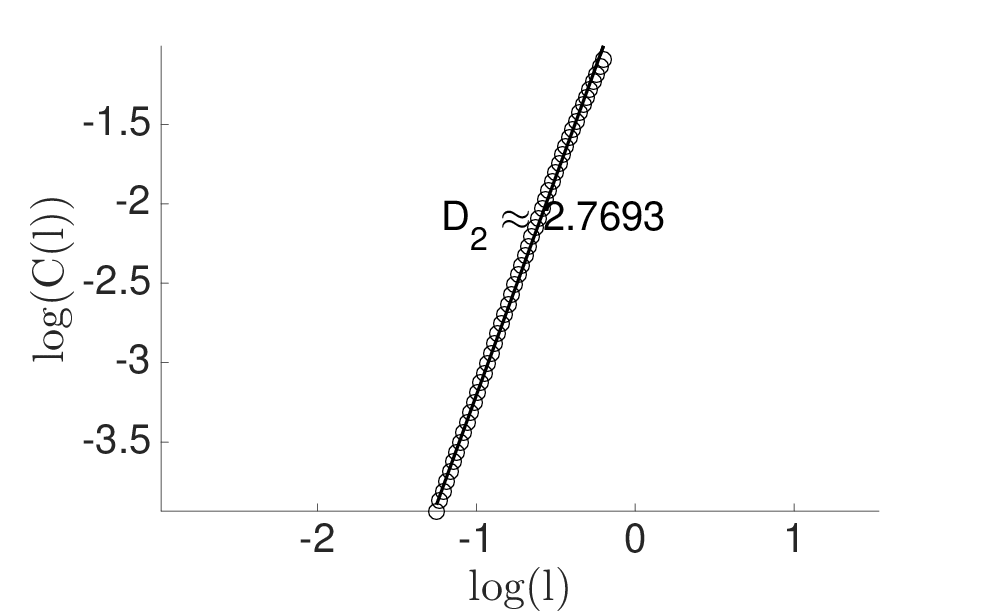}
            \caption{}
            \end{subfigure}
            \caption{Result for heart rate($x_n$) vs breathing rate($y_n$) data. Top raw is the scatter plot of the data and second raw represent the dimension of the data.  }\label{HBdata}
        \end{figure}

Now, moving beyond bench-marking with synthetic data, we will contrast the %three
two measures of information flow in a real world experimental data set. 
 Consider heart rate ($x_n$ ) vs breathing rate ($y_n$) data (Fig.~ \ref{HBdata}) as published in \cite{HrateData1,HrateData2}, consisting of $5000$ samples. Correlation dimension of the data $X$ is $D_2(X)= 1.00$, and  $D_2(X,X')=1.8319>D_2(X)$. Therefore, $X'=X_{n+1}$ depends not only $x$ but also on an extra variable (Thm.~\ref{dimTheorem}). Also correlation dimension of the data $(X,Y)$ and  $(X,Y,X')$ is computed $D_2(X,Y)=1.9801$ and $D_2(X,Y,X')=2.7693>D_2(X,Y)$ respectively. We conclude that $X'$ depends on extra variable(s) other that $(x,y)$ (Thm.~ \ref{dimTheorem}) and the correlation dimension geometric information flow,  $GeoC_{y\to x}=0.0427$, is computed by Eqs.~(\ref{geocauC})-(\ref{geocau2}). 
 Therefore, this suggests the conclusion that there is a causal inference from breathing rate to heart rate. Since breathing rate and heart rate share the same units, the quantity measured by geometric information flow can be described without normalizing. Transfer entropy as estimated by the KSG method (\cite{PhysRevE.69.066138}) with parameter $k=30$ is $T_{y\to x}=0.0485$, interestingly relatively close to the $GeoC$ value. 
 In summary, %all three 
 both measures for causality ($GeoC,\ T$) are either zero or positive together. It follows that there exists a causal inference.

 \begin{table}[H]
\caption{Heart rate vs breathing rate data.  Contrasting geometric information flow versus transfer entropy in breath rate to heart rate.}
\centering 
\begin{tabular}{ c c} 
\hline\hline 
$GeoC_{y\to x}$ & $T_{y\to x}$  \\ [0.5ex]
\hline 
 0.0427&0.0485  \\
[1ex] 
\hline 
\end{tabular}
\label{table:hrateVbreate} 
\end{table}

\section{Conclusion}  

 We have developed here a geometric interpretation of information flow as a causal inference as usually measured by a positive transfer entropy. $T_{y\rightarrow x}$.  Our interpretation relates the dimensionality of an underlying manifold as projected into the outcome space  summarizes the information flow.  Further, the analysis behind our interpretation involves standard Pinsker's inequality that estimates entropy in terms of total variation, and through this method we can interpret the production of information flow in terms of details of the derivatives describing relative orientation of the manifolds describing inputs and outputs (under certain simple assumptions).
 
A geometric description of causality allows for new and efficient  computational methods for causality inference. Furthermore, this geometric prospective provide different view of the problem and facilitate the richer understanding that complement the probabilistic descriptions. Causal inference is weaved strongly throughout many  fields and the use of transfer entropy as been a popular black box tool for this endeavor. Our method can be used to reveal more details of the underling geometry of the data-set and provide clear view of the causal inference. In addition, one can use hybrid method of this geometric aspect and existing other method in their applications.
 
 We provided a theoretical explanation (part I: Mathematical proof of the geometric view of the problem)  and numerical evidence (part 2: a data-driven approach for mathematical framework) of geometric view for the causal inference. Our experiments are based on synthetic (toy problems) and practical data. In the case of synthetic data, the underlining dynamics of the data and the actual solution to the problem is known. Each of these toy problems we consider a lot of cases by setting a few parameters. Our newly designed geometric approach can successfully capture these cases.  One major problem may be if data describes a chaotic attractor. We prove theoretically (theorem \ref{dimTheorem}) and experimentally (by H\'{e}non map example: in this toy problem we also know actual causality) that correlation dimension serves to overcome this issue. Furthermore, we present a practical example based on heart rate vs breathing rate variability, which was already shown to have positive transfer entropy, and here we relate this to show positive geometric causality.
 
 Further, we have pointed out that transfer entropy has analytic convergence issues when future data ($X'$) is exactly a function of current input data ($X,Y$) versus more generally $(X,Y,X')$. Therefore, referring to how the geometry of the data can be used to identify the causation of the time series data, we develop a new causality measurement based on a fractal measurement comparing inputs and outputs.  Specifically,  the correlation dimension is a useful and efficient way to define what we call  correlation dimensional geometric information flow, $GeoC_{y\rightarrow x}$. The $GeoC_{y\rightarrow x}$ offers a strongly geometric interpretable result as a global picture of the information flow. We demonstrate the natural benefits of $GeoC_{y\rightarrow x}$ versus $T_{y\rightarrow x}$,  in several synthetic examples where we can specifically control the geometric details, and then with a physiological example using heart and breathing data.

%%%%%%%%%%%%%%%%%%%%%%%%%%%%%%%%%%%%%%%%%%
\vspace{6pt} 

%%%%%%%%%%%%%%%%%%%%%%%%%%%%%%%%%%%%%%%%%%
%% optional
%\supplementary{The following are available online at \linksupplementary{s1}, Figure S1: title, Table S1: title, Video S1: title.}


%%%%%%%%%%%%%%%%%%%%%%%%%%%%%%%%%%%%%%%%%%
\authorcontributions{Conceptualization, Sudam Surasinghe and Erik M. Bollt; Data curation, Sudam Surasinghe and Erik M. Bollt; Formal analysis, Sudam Surasinghe and Erik M. Bollt; Funding acquisition, Erik M. Bollt; Methodology, Sudam Surasinghe and Erik M. Bollt; Project administration, Erik M. Bollt; Resources, Erik M. Bollt; Software, Sudam Surasinghe and Erik M. Bollt; Supervision, Erik M. Bollt; Validation, Sudam Surasinghe and Erik M. Bollt; Visualization, Sudam Surasinghe and Erik M. Bollt; Writing – original draft, Sudam Surasinghe and Erik M. Bollt; Writing – review \& editing, Sudam Surasinghe and Erik M. Bollt.}

%%%%%%%%%%%%%%%%%%%%%%%%%%%%%%%%%%%%%%%%%%
\funding{EB gratefully acknowledges funding from the Army Research Office W911NF16-1-0081 (Dr Samuel Stanton) as well as from DARPA.}

%%%%%%%%%%%%%%%%%%%%%%%%%%%%%%%%%%%%%%%%%%
\acknowledgments{ We would like to thank Professor Ioannis Kevrekidis for his generous time, feedback and interest regarding this project.}

%%%%%%%%%%%%%%%%%%%%%%%%%%%%%%%%%%%%%%%%%%
%\conflictsofinterest{Declare conflicts of interest or state ``The authors declare no conflict of interest.'' Authors must identify and declare any personal circumstances or interest that may be perceived as inappropriately influencing the representation or interpretation of reported research results. Any role of the funders in the design of the study; in the collection, analyses or interpretation of data; in the writing of the manuscript, or in the decision to publish the results must be declared in this section. If there is no role, please state ``The funders had no role in the design of the study; in the collection, analyses, or interpretation of data; in the writing of the manuscript, or in the decision to publish the results''.} 

%%%%%%%%%%%%%%%%%%%%%%%%%%%%%%%%%%%%%%%%%%
%% optional
%\abbreviations{The following abbreviations are used in this manuscript:\\

%\noindent 
%\begin{tabular}{@{}ll}
%MDPI & Multidisciplinary Digital Publishing Institute\\
%DOAJ & Directory of open access journals\\
%TLA & Three letter acronym\\
%LD & linear dichroism
%\end{tabular}}

%%%%%%%%%%%%%%%%%%%%%%%%%%%%%%%%%%%%%%%%%%
%% optional
\appendixtitles{yes} %Leave argument "no" if all appendix headings stay EMPTY (then no dot is printed after "Appendix A"). If the appendix sections contain a heading then change the argument to "yes".
\appendix
\section{}
\unskip

\subsection{On the Asymmetric Spaces Transfer Operators}\label{dis}
In this section we prove Theorem \ref{ThmASTO} concerning a transfer operator for smooth transformations  between sets of perhaps dissimilar dimensionality.  In general, the marginal probability density can be found by integrating (or summation in the case of a discrete random variable) to marginalize the joint probability densities. When $x'=f(x,y)$, the  joint density $(x,y,x')$  is non-zero only at points on $x'=f(x,y)$. Therefore $\rho(x')=\sum_{(x,y):x'=f(x,y)}\rho(x,y,x')$ and notice that $\rho(x,y,x')=\rho(x'|x,y)\rho(x,y)$ (By Bayes theorem). Hence $\rho(x')=\sum_{(x,y):x'=f(x,y)}\rho(x'|x,y)\rho(x,y)$ and we only need to show the following claims.  We will discuss this by two cases. First we consider $x'=f(x)$ and then we consider more general case $x=f(x,y)$. In higher dimensions we can consider similar scenarios of input and output variables, and correspondingly the trapezoidal bounding regions would need to be specified in which we can analytically control the variables.

\begin{Proposition}[Claim]
Let $X\in \mathbb{R}$ be a random variable with probability density function $\rho(x)$. Suppose $\rho(x), \rho(.|x)$ are Radon–Nikodym derivatives (of induced measure with respect to some base measure $\mu$) which is bounded above and bounded away from zero.  Also let $x'=f(x)$ for some function $f\in C^1(\mathbb{R})$. Then \[\rho(x'|X=x_0)= \lim_{\epsilon \to 0}d_{\epsilon}(x'-f(x_0))\] where $d_{\epsilon}(x'-f(x_0))=\begin{cases} \frac{1}{2\epsilon |f'(x_0)|} &; |x'-f(x_0)|<\epsilon|f'(x_0)| \\ 0 & ;\  \textit{ otherwise} \end{cases}.$
\end{Proposition}
\begin{proof}
Let $1>>\epsilon>0$ and $x\in I_{\epsilon}=(x_0-\epsilon,x_0+\epsilon)$. Since $\rho$ is  a Radon–Nikodym derivative with bounded above and bounded away from zero, $\rho(I_{\epsilon}) = \int_{I_{\epsilon}} \frac{d\rho}{d\mu} d\mu \ge  \frac{m}{2\epsilon}$ where $m$ is the infimum of the Radon–Nikodym derivative. Similarly $\rho(I_{\epsilon}) \le \frac{M}{2\epsilon}$ where $M$ is the supremum of the Radon–Nikodym derivative. Also $|x'-f(x_0)|\approx |f'(x_0)||x-x_0|$ for $x\in I_{\epsilon}$. Therefore, $x'\in (f(x_0)-\epsilon |f'(x_0)|,f(x_0)+\epsilon |f'(x_0)|)=I'_{\epsilon}$ when $x\in I_{\epsilon}$. Hence $\rho(x'|x\in I_{\epsilon})=\rho(x'\in I'_{\epsilon})$ and $\frac{m}{2\epsilon |f'(x_0)|} \le \rho(x'|x\in I_{\epsilon})\le \frac{M}{2\epsilon |f'(x_0)|}$. Therefore $\rho(x'|X=x_0)= \lim_{\epsilon \to 0}d_{\epsilon}(x'-f(x_0))$.
\end{proof}
\begin{Proposition}[Claim] 2
Let $X,Y\in \mathbb{R}$ be random variables with joint probability density function $\rho(x,y)$. Suppose $\rho(x,y)$ and $\rho(.|x,y)$ are Radon–Nikodym derivatives (of induced measure with respect to some base measure $\mu$) which is bounded above and bounded away from zero. Also let $x'=f(x,y)\in \mathbb{R}$ for some function $f\in C^1(\mathbb{R})$. Then \[\rho(x'|X=x_0, Y=y_0)= \lim_{\epsilon \to 0}d_{\epsilon}(x'-f(x_0,y_0))\] where $d_{\epsilon}(x'-f(x_0,y_0))=\begin{cases} \frac{1}{2\epsilon (|f_x(x_0,y_0)|+|f_y(x_0,y_0)|)} &; |x'-f(x_0,y_0)|<\epsilon(|f_x(x_0,y_0)|+|f_y(x_0,y_0)|) \\ 0 & ;\  \textit{ otherwise} \end{cases}.$
\end{Proposition}
\begin{proof}
Let $1>>\epsilon>0$ and $A_{\epsilon}=\{(x,y)|x\in (x_0-\epsilon,x_0+\epsilon), y\in (y_0-\epsilon,y_0+\epsilon)$ . Since $\rho$ is   a Radon–Nikodym derivative with bounded above and bounded away from zero, $\rho(A_{\epsilon}) = \int_{A_{\epsilon}} \frac{d\rho}{d\mu} d\mu \ge  \frac{m}{4\epsilon^2}$ where $m$ is the infimum of the Radon–Nikodym derivative. Similarly, $\rho(A_{\epsilon}) \le \frac{M}{4\epsilon^2}$ where $M$ is the supremum of the Radon–Nikodym derivative. Also $|x'-f(x_0,y_0)|\approx |f_x(x_0,y_0)||x-x_0|+|f_y(x_0,y_0)||y-y_0|$ for $(x,y)\in A_{\epsilon}$ . Therefore, $x'\in (f(x_0,y_0)-\epsilon (|f_x(x_0,y_0)|+|f_y(x_0,y_0)|),f(x_0,y_0)+\epsilon (|f_x(x_0,y_0)|+|f_y(x_0,y_0)|))=I'_{\epsilon}$ when $(x,y)\in A_{\epsilon}$. Hence $\rho(x'|(x,y)\in A_{\epsilon})=\rho(x'\in I'_{\epsilon})$ and $\frac{m}{2\epsilon (|f_x(x_0,y_0)|+|f_y(x_0,y_0)|)} \le \rho(x'|x\in I_{\epsilon})\le \frac{M}{2\epsilon (|f_x(x_0,y_0)|+|f_y(x_0,y_0)|)}$. Therefore $\rho(x'|X=x_0, Y=y_0)= \lim_{\epsilon \to 0}d_{\epsilon}(x'-f(x_0,y_0))$.

\end{proof}
If $f$ only depends on $x$, then the partial derivative of $f$ with respect to $y$ is equal to zero and which leads to the same result as clam 1. 
%\section{}
%All appendix sections must be cited in the main text. In the appendixes, Figures, Tables, etc. should be labeled starting with `A', e.g., Figure A1, Figure A2, etc. 


%=====================================
% References, variant A: internal bibliography
%=====================================
\reftitle{References}
% The following MDPI journals use author-date citation: Arts, Econometrics, Economies, Genealogy, Humanities, IJFS, JRFM, Laws, Religions, Risks, Social Sciences. For those journals, please follow the formatting guidelines on http://www.mdpi.com/authors/references
% To cite two works by the same author: \citeauthor{ref-journal-1a} (\citeyear{ref-journal-1a}, \citeyear{ref-journal-1b}). This produces: Whittaker (1967, 1975)
% To cite two works by the same author with specific pages: \citeauthor{ref-journal-3a} (\citeyear{ref-journal-3a}, p. 328; \citeyear{ref-journal-3b}, p.475). This produces: Wong (1999, p. 328; 2000, p. 475)

%=====================================
% References, variant B: external bibliography
%=====================================
\externalbibliography{yes}
\bibliography{bibit}

%%%%%%%%%%%%%%%%%%%%%%%%%%%%%%%%%%%%%%%%%%
%% optional
%\sampleavailability{Samples of the compounds ...... are available from the authors.}

%% for journal Sci
%\reviewreports{\\
%Reviewer 1 comments and authors’ response\\
%Reviewer 2 comments and authors’ response\\
%Reviewer 3 comments and authors’ response
%}

%%%%%%%%%%%%%%%%%%%%%%%%%%%%%%%%%%%%%%%%%%
\end{document}